\documentclass[lineno]{jfm}

\usepackage{graphicx}
\usepackage{newtxtext}
\usepackage{newtxmath}
\usepackage{natbib}
\usepackage{hyperref}
\hypersetup{
    colorlinks = true,
    urlcolor   = blue,
    citecolor  = black,
}

\newcommand{\RomanNumeralCaps}[1]
\linenumbers

\newcommand\Pra{\mbox{\textit{Pr}}}
\newcommand\Sch{\mbox{\textit{Sc}}}
\newcommand\Ray{\mbox{\textit{Ra}}}
\newcommand\Nus{\mbox{\textit{Nu}}}
\newcommand\Lew{\mbox{\textit{Le}}}

\title{On the wall-bounded model of fingering double diffusive convection}

\author{Junyi Li\aff{1,2}
        \and Yantao Yang\aff{1,2}\corresp{\email{yantao.yang@pku.edu.cn}}}

\affiliation{
\aff{1} State Key Laboratory for Turbulence and Complex Systems, and Department of Mechanics and Engineering Science, College of Engineering, Peking University, Beijing 100871, P. R. China \\
\aff{2} Joint Laboratory of Marine Hydrodynamics and Ocean Engineering, Pilot National Laboratory for Marine Science and Technology (Qingdao), Shandong 266299, P. R. China
}

\begin{document}

\maketitle

\begin{abstract}
Fingering double diffusive convection with real seawater properties is studied by two-dimensional direct numerical simulations for the wall-bounded domain and compared with the results for fully periodic domain. For fixed unstable salinity difference between two horizontal plates, dominant flow structures change from convection rolls to salt fingers as the stable temperature difference increases. Meanwhile the bulk density ratio calculated by the mean scalar gradients exceeds unity. When the bulk thermal Rayleigh number, which is defined by the mean temperature gradient in the bulk and the domain height, is larger than $10^7$, the characteristic height of salt fingers is much smaller than the domain height and the flow enters the free salt-finger regime. In this regime the transport properties agree quantitatively with those obtained in the fully periodic domain (e.g. Traxler et al. \emph{J. Fluid Mech.}, \textbf{677}, 530-553, 2011). The salt-finger bulk does not spontaneously break into multi-layer staircases probably due to the fact that solid boundary prevents the development of large-scale secondary instabilities. For the limited range of density ratio at the highest salinity Rayleigh number considered here, the multi-layer state is directly established from the initial condition with uniform salinity distribution and vertically linear temperature distribution.
\end{abstract}

\begin{keywords}

\end{keywords}

\section{Introduction}\label{sec:intro} 

When fluid density is determined by two scalar components with different molecular diffusivities, double diffusive convection (DDC) may occur if the stratifications of scalar components are in a suitable configuration. In the Ocean, DDC is omnipresent as the vertical gradients of temperature and salinity favour DDC instability in many regions~\citep{you2002,schmitt2005,shibley2017,durante2019}. Note that temperature diffuses about 100 times faster than salt, and very rich dynamics can be excited due to this huge difference in diffusivity. In the (sub-)tropic ocean, usually both temperature and salinity decrease with depth in the upper water, where DDC happens mainly in the fingering regime~\citep{you2002,schmitt1994}. In fingering DDC (FDDC) the salinity gradient drives the fluid motion, while the temperature gradient stabilizes the flow. FDDC can occur when the overall density is stably stratified~\citep{stern1960}, thus greatly extends the environments for convection motions, and plays an important and unique role in oceanic mixing~\citep{schmitt2005}.

Numerous efforts have been made to understand the physical mechanisms and transport properties of FDDC. Reviews of early observations, experiments, simulations, and theoretical models can be found in~\citet{schmitt2003,yoshida2003,kunze2003} and the book of~\citet{radko2013}. Since FDDC represents a small-scale phenomenon in the Ocean, it is very challenging to obtain detailed information in field measurements. Experiments are also very difficult in the sense that two scalar components have to be controlled and measured simultaneously. In numerical simulations, though, it is very convenient to precisely control the flow conditions and acquire all the information of the flow fields. 

One major challenge in simulations is how to deal with the very small molecular diffusivity of salinity, which is usually three orders of magnitude smaller than viscosity. Scalar with small diffusivity requires very fine resolution to be fully resolved. In addition, the salt-sugar system is often used in the laboratory experiments~\citep{krishnamurti2003}, in which the ratio of diffusivities between two scalars decreases to about 3. In many numerical studies, therefore, salinity is replaced by a scalar with larger diffusivity~\citep[see for example][]{stellmach2011,paparella2012}. Another technique is the multiple-resolution method as developed in our previous work~\citep{multigrid2015}, in which salinity is solved on a refined mesh. With the help of this efficient method, very large control parameters have been achieved for the same fluid properties as seawater in fully three-dimensional (3D) simulations~\citep{ddcjfm2016}.

Different configurations of flow domain were employed in the existing numerical investigations of FDDC. One type is the so-called ``run-down'' configuration, in which two homogeneous layers are separated by an interface~\citep{sreenivas2009} and the system is isolated without any heat or salt exchange with the outside. The top layer has both higher temperature and salinity so that salt fingers develop around the initial interface. This configuration is identical to many early experimental setup, such as~\cite{turner1967,linden1973,schmitt1979}. Since the total potential energy is fixed by the initial field, the system undergoes continuous transition until the available energy is completely consumed, i.e. the flow cannot reach a statistically steady state.

In order to maintain a statistically steady state, a constant driving force should be introduced. Two typical choices have been utilized. The first one employs constant background temperature and salinity gradients and simulates the temperature and salinity deviated from this background field. Fully periodic domain can then be used. When the ratio between the background temperature and salinity gradients exceeds unity a little, FDDC can be realized efficiently by standard pseudo-spectral scheme~\citep{traxler2011,stellmach2011}. Another choice is the wall-bounded model which is commonly used in thermal convection~\citep{ahlers2009}. In this configuration a fluid layer is bounded from top and bottom by two parallel plates which usually have constant temperature and salinity. Therefore, constant differences in temperature and salinity are maintained across the layer. Wall-bounded FDDC has been investigated in both experiments and numerical simulations~\citep{radko2000,krishnamurti2003,hage2010,kellner2014,ddcjfm2015,ddcjfm2016}.

In the wall-bounded model boundary layers develop adjacent to the two plates in momentum, temperature and salinity fields. The appearance of boundary layers and their interaction with the salt fingers in the bulk inevitably affect the flow dynamics and transport behaviours~\citep{radko2000,ddcjfm2016}. For fixed salinity difference between two plates, the flow morphology can shift from wide convection rolls at small temperature difference to slender salt fingers at large temperature difference~\citep{hage2010,kellner2014,ddcpnas2016}. Our recent simulations of wall-bounded FDDC further reveal that multiple equilibrate states can be established for the exactly same global flow parameters~\citep{ddcpnas2020}. In the same work we also show that in the wall-bounded model with very high salinity and temperature differences, different initial distributions of temperature and salinity lead to staircases with different layer configurations. Therefore, the wall-bounded FDDC provides a unique system to study the dynamics and evolution of fully developed FDDC staircases.  

One concern about the wall-bounded FDDC model is the existence of the solid plates, which are not presented in the oceanic FDDC. The free-slip condition can be used to eliminate the viscous drag along the two plates, but the effects of non-penetration condition still exit. Our previous study indeed shows that wall-bounded FDDC with free-slip and no-slip boundary exhibit very similar behaviours in flow structures and transport properties~\citep{ddcprl2016}. In the triply periodic domain, the domain size needs to be large enough to remove the numerical constraints on finger length scales~\citep{traxler2011}. However, if the domain is too large, secondary large-scale instabilities can develop and drive the system away from pure finger state. It is also worthy to mention that the so-called ``elevator modes'' which grow exponentially in the triply periodic Rayleigh-B\'{e}nard (RB) convection~\citep{calzavarini2006} are exactly the tall-finger modes in triply periodic FDDC~\citep{schmitt1979,radko2013}. Apparently, such elevator modes occupying the whole domain height are prevented by the two solid plates in wall-bounded FDDC. 

Therefore, the aims of this study are to clarify the effects of solid boundary in wall-bounded FDDC and to establish the correspondence between the wall-bounded model and triply periodic model for FDDC. We will conduct systematic simulations of FDDC with the fluid properties same as seawater, and identify the parameter regime where the influences of solid boundary are negligible. The rest of paper is organized as follows. In \S~\ref{sec:method} we describe the governing equations and numerical methods. We then discuss the flow structures and transport properties in \S~\ref{sec:struc} and \S~\ref{sec:trans}, respectively. Conclusions are given in \S~\ref{sec:conclusion}.

\section{Governing Equations and Numerical Methods}\label{sec:method} 

We consider a fluid layer bounded by two parallel plates from top and bottom. The two plates are perpendicular to the gravity and separated by a height $H$. We employ a linear equation of state as $\rho^*=\rho^*_0(1-\beta_T \theta^* +\beta_S s^*)$. Here $\rho^*$ is density, with the subscript ``0'' denoting the value at the reference state. $\theta^*$ and $s^*$ are the temperature and salinity with respect to the corresponding reference values. $\beta_T$ is the thermal expansion coefficient, and $\beta_S$ is the linear coefficient related to the salinity, respectively. Hereafter the asterisk denotes the dimensional quantity. Then, under the Oberbeck-Boussinesq approximation, the governing equations read
\begin{eqnarray}
  \partial_t \boldsymbol{u}^* + \boldsymbol{u}^*\bcdot\bnabla\boldsymbol{u}^* &=& -\bnabla p^* + \nu \bnabla^2\boldsymbol{u}^* + g(\beta_T \theta^* - \beta_S s^*)\boldsymbol{e}_z, \label{eq:u} \\
  \partial_t \theta^* + \boldsymbol{u}^*\bcdot\bnabla \theta^* &=& \kappa_T \nabla^2\theta^*, \label{eq:t} \\
  \partial_t s^* + \boldsymbol{u}^*\bcdot\bnabla s^* &=& \kappa_S \nabla^2 s^*, \label{eq:s}\\
  \bnabla\bcdot\boldsymbol{u}^* &=& 0. \label{eq:c}
\end{eqnarray}
Here, $\boldsymbol{u}$ is velocity, $p$ is pressure, $g$ is the gravitational acceleration, $\nu$ is viscosity, and $\kappa$ is molecular diffusivity, respectively. In equation~\eqref{eq:u} density has been absorbed into pressure. $\boldsymbol{e}_z$ is the unit vector in the vertical $z$-direction. The two plates are set as non-slip walls with constant temperature and salinity. The top plate has higher temperature and salinity so that the system is in the FDDC regime. In the horizontal directions the periodic boundary conditions are applied. The boundary conditions then read
\begin{subeqnarray}
  \boldsymbol{u}^*=\boldsymbol{0},~s^*=\Delta_S,~\theta^*=\Delta_T, & \quad \mbox{at} &~z^*/H=1, \\
  \boldsymbol{u}^*=\boldsymbol{0},~s^*=0,~\theta^*=0, \quad~ & \quad \mbox{at} &~z^*/H=0. 
\end{subeqnarray}
Here the fluid at bottom plate is chosen as the reference state. $\Delta_T$ and $\Delta_S$ are the constant temperature and salinity differences between the two plates, respectively.

The governing equations are non-dimensionalized by the height $H$, the constant temperature and salinity differences $\Delta_T$ and $\Delta_S$ between the two plates, and the free-fall velocity $\sqrt{\,gH\beta_S\Delta_S\,}$, respectively. The control parameters include the Prandtl number, the Schmidt number, and two Rayleigh numbers, which are defined respectively as
\begin{equation}
 \Pra = \frac{\nu}{\kappa_T}, \quad
 \Sch = \frac{\nu}{\kappa_S}, \quad
 \Ray_T = \frac{g\beta_T\Delta_T H^3}{\nu\kappa_T}, \quad
 \Ray_S = \frac{g\beta_S\Delta_S H^3}{\nu\kappa_S}.
\end{equation}
Throughout this study we fix $\Pra=7$ and $\Sch=700$, which are the typical values for temperature and salinity in the Ocean. The Lewis number, i.e.~the ratio between the two diffusivities, is then $\Lew=\Sch/\Pra=100$. The relative strength of the temperature difference to the salinity difference can be measured by the density ratio as
\begin{equation}
  \Lambda = \frac{\beta_T\,\Delta_T}{\beta_S\,\Delta_S} 
  = \frac{\Sch\,\Ray_T}{\Pra\,\Ray_S} = \frac{\Lew\,\Ray_T}{\Ray_S}.
\end{equation}
Then the non-dimensional governing equations are
\begin{eqnarray}
  \partial_t \boldsymbol{u} + \boldsymbol{u}\bcdot\bnabla\boldsymbol{u} &=& 
     -\bnabla p + \Sch^{1/2}\Ray^{-1/2}_S\,\bnabla^2\boldsymbol{u} + (\Lambda \theta - s)\boldsymbol{e}_z, \label{eq:und} \\
  \partial_t \theta + \boldsymbol{u}\bcdot\bnabla \theta &=& Sc^{1/2}\Ray_S^{-1/2}\Pra^{-1}\,\nabla^2\theta, \label{eq:tnd} \\
  \partial_t s + \boldsymbol{u}\bcdot\bnabla s &=& \Sch^{-1/2}\Ray_S^{-1/2}\,\nabla^2 s, \label{eq:snd}\\
  \bnabla\bcdot\boldsymbol{u} &=& 0. \label{eq:cnd}
\end{eqnarray}
with the boundary conditions
\begin{subeqnarray}
  \boldsymbol{u}=\boldsymbol{0},~s=1,~\theta=1, & \quad \mbox{at} &~z=1, \\
  \boldsymbol{u}=\boldsymbol{0},~s=0,~\theta=0, & \quad \mbox{at} &~z=0. 
\end{subeqnarray}

\begin{figure}
	\centerline{\includegraphics[width=0.8\textwidth]{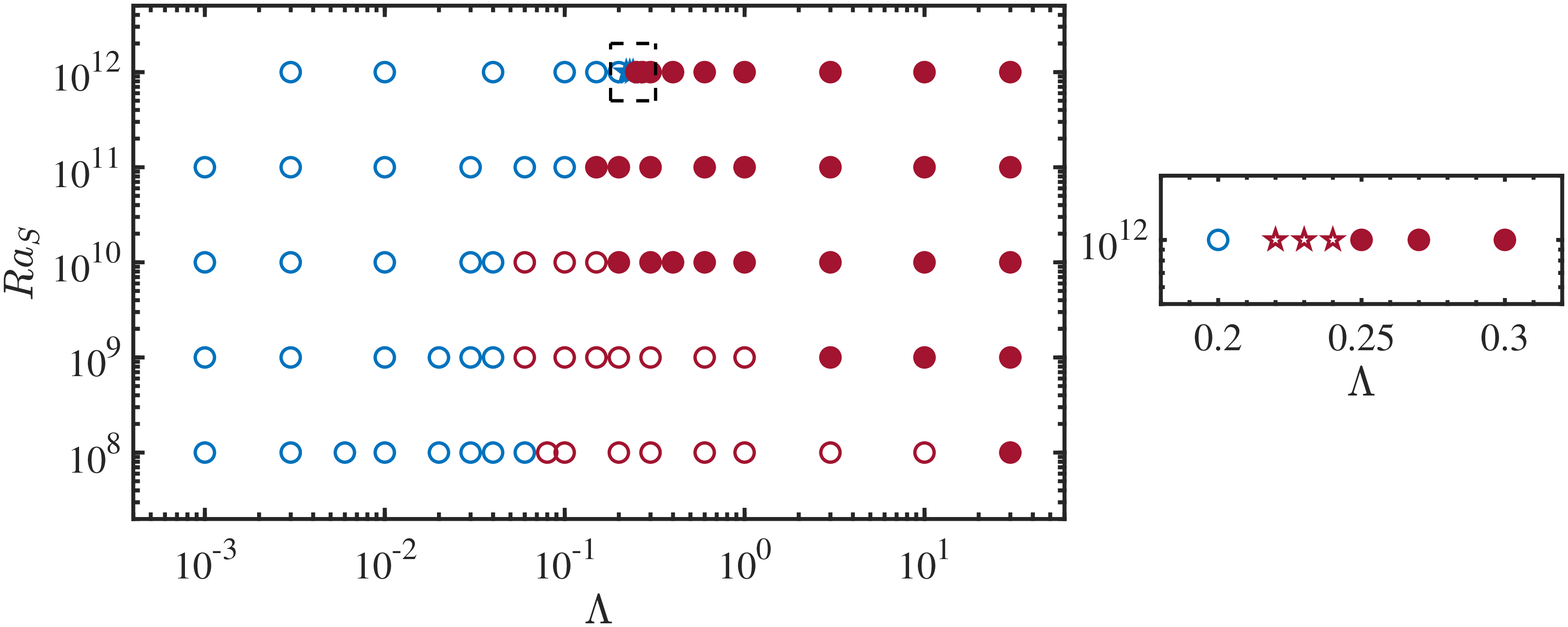}}
	\caption{ The parameter space on the $\Lambda$-$Ra_S$ plane explored in the current study. The zoom-in plot of the dashed box are shown in the right panel. The blue and red circles mark the cases of convection-roll type and salt-finger type, respectively. The red open circles denote the confined-salt-finger regime, while the solid circles denote the free-salt-finger regime, as defined below in figure~\ref{fig:zscale}. The star symbols at $\Ray_S=10^{12}$ indicate the cases of the multi-layer staircase state, while all other cases are at the single finger-layer state.}
	\label{fig:params}
\end{figure}

The non-dimensional governing equations (\ref{eq:und})-(\ref{eq:cnd}) are numerically solved by using our in-house code, which employs the finite difference scheme and a fractional time-step method~\citep{multigrid2015}. Especially, the code utilizes a dual-resolution technique to deal with the salinity field which has a very high Schmidt number of 700. A base mesh is used for the momentum and temperature fields, and a refined mesh for the salinity field, respectively. A locally mass conserved interpolation method is developed to construct the velocity field at the refined mesh from that at the base mesh. The code has been extensively applied to FDDC in our previous works~\citep{ddcjfm2015,ddcjfm2016,ddcprl2016}, and validated by a one-to-one comparison with experiments~\citep{ddcjfm2015}. Still, fully three-dimensional (3D) simulations with $\Pra=7$ and $\Sch=700$ are very challenging for a systematic study. In the present work we conduct two-dimensional (2D) simulations for a wide range of control parameters, see the phase diagram shown in figure~\ref{fig:params}. It has been shown that for FDDC both 2D and 3D numerical results exhibit very similar behaviours~\citep{ddcpnas2020}. Therefore, 2D simulations can still provide valuable insights into the physics of FDDC. 

In the wall-bounded model the flow morphology can shift from wide convection rolls to slender fingers~\citep{kellner2014,ddcpnas2016}. In order to consistently investigate all the flow morphologies, the salinity Rayleigh number $\Ray_S$, which measures the strength of driving force, is chosen as one primary global control parameter. We simulate five different salinity Rayleigh numbers ranging between $10^{8}$ and $10^{12}$. The global density ratio $\Lambda$ is then systematically changed for fixed strength of driving force. For the four lower Rayleigh numbers we systematically increase $\Lambda$ from $10^{-3}$ up to $30$. While for the highest $Ra_S=10^{12}$ the smallest density ratio is $\Lambda=0.004$ to save simulation cost. Note that we choose $\Lambda$ starting at the value far below unity, since the salt fingers can develop in the bulk of wall-bounded domain even when the overall density is unstably stratified~\citep{hage2010,schmitt2011}, and the transition from wide convection rolls to slender salt fingers happens at strongly unstable density stratification~\citep{kellner2014,ddcpnas2016}. The details about the numerics and the global responses are summarized in the Appendix.

\section{On the flow structures in the bulk}\label{sec:struc} 

\subsection{The initial development of flows}

We first discuss the choice of initial conditions and the temporal evolution of the flow fields afterwords. For all the simulations the initial temperature field is a linear distribution between the two plates, while the initial salinity field is uniform and equals to the mean of the values at two plates, respectively. The fluid is initially at rest. Small perturbations are added to trigger the flow motions. These initial conditions are the same as in the experiments of~\cite{hage2010} and our previous simulations~\citep{ddcjfm2015,ddcjfm2016}. Our previous work reveals that for a fixed $\Lambda$ when $\Ray_S$ is above some critical value, multiple equilibrate staircase states can be established with the same global control parameters~\citep{ddcpnas2020}. However, once a single finger-layer state is achieved and occupies the entire bulk, it is stable even when $\Ray_S$ is larger than the corresponding critical value.

For the salinity Rayleigh numbers considered here, most of cases can reach the single finger-layer state from the above initial fields except for a couple of cases at highest $\Ray_S=10^{12}$. If the final state at a given set of control parameters is a single finger layer, flows starting from different initial conditions will reach the same final state through different evolution processes. To demonstrate this, we run an extra case for $\Ray_S=10^{10}$ and $\Lambda=0.1$ with both scalar components having a vertically linear distribution, which we refer to as the linear initial condition. The one used in all the simulations is referred to as the mixed initial condition. To quantitatively illustrate the flow evolution, we define the instantaneous bulk density ratio and Reynolds number as
\begin{equation}
\Lambda_b^{in}=\Lambda \frac{ T_z^{in}} {S_z^{in}}, \quad
\Rey^{in} = \frac{U^{in}_{rms} H}{\nu}. 
\end{equation}
Here $T_z^{in}$ and $S_z^{in}$ denote the dimensionless instantaneous vertical gradients of the horizontal averaged temperature and salinity profiles, respectively. The two gradients are calculated by the linear fitting of $\langle \theta \rangle_h$ and $\langle s \rangle_h$ over the range of $0.25\le z \le 0.75$. From now on, the bracket $\langle \rangle_h$ denotes the horizontal averaged value. $U^{in}_{rms}$ is the dimensional instantaneous root-mean-square (rms) value of the magnitude of velocity vector, which is computed over the entire domain.

\begin{figure}
	\centerline{\includegraphics[width=1\textwidth]{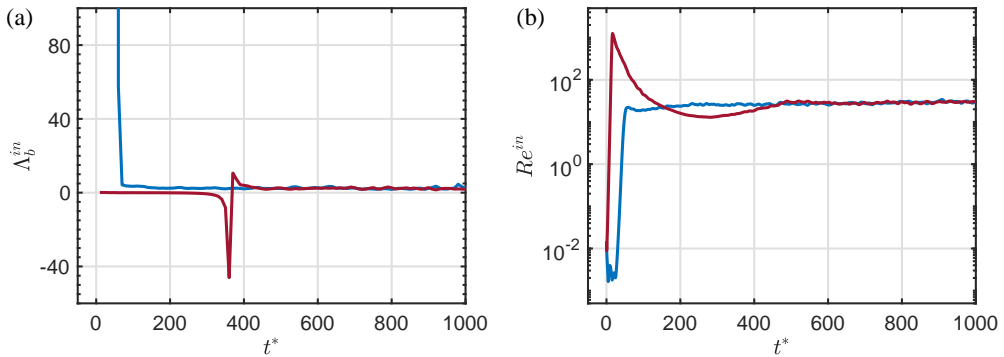}}
	\caption{The comparison of the temporal evolution of (a) instantaneous bulk density ratio and (b) Reynolds number for the two cases starting from the mixed initial condition (blue lines) and the linear initial condition (red lines). The global control parameters are $Ra_S=10^{10}$ and $\Lambda=0.1$.}
	\label{fig:timehist1e10}
\end{figure}

\begin{figure}
	\centerline{\includegraphics[width=\textwidth]{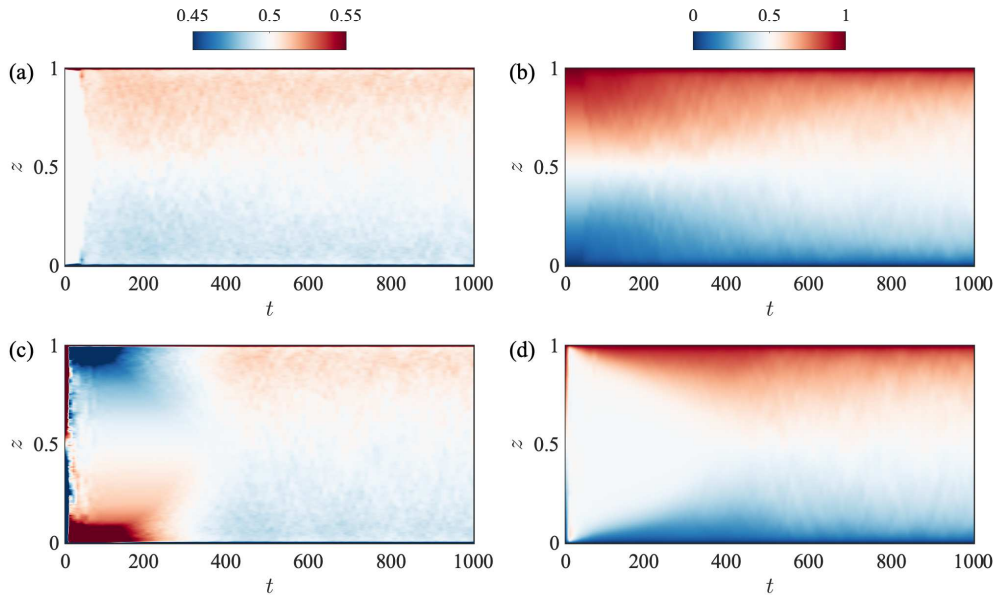}}
	\caption{The temporal evolution of the horizontally averaged scalar profiles staring from different initial conditions for $\Ray_S=10^{10}$ and $\Lambda=0.1$. Panels a and b show the salinity and temperature profiles for the case with mixed initial condition, respectively. Panels c and d show the same quantities for the case with linear initial condition.}
	\label{fig:sthist1e10}
\end{figure}

Figure~\ref{fig:timehist1e10} shows the time evolution of $\Lambda_b^{in}$ and $Re^{in}$ for the two cases with mixed and linear initial conditions. And the salinity fields at two different times are compared in figure~\ref{fig:sthist1e10}. For the case with mixed initial condition, $\Lambda^{in}_b$ is very large at the beginning since $S^{in}_z$ is close to zero. As plumes start to grow from the top and bottom boundaries where the salinity field is strongly unstable, $\Rey^{in}$ increases rapidly and $\Lambda^{in}_b$ decreases towards the value of final statistically steady state. From the evolution of mean temperature and salinity profiles shown in figures~\ref{fig:sthist1e10}a and b, one can see that nearly linear profiles directly build up in the bulk as buoyancy-driven motions develop with time. Meanwhile, the case with linear initial condition undergoes a totally different route of initial development. Since for $\Lambda=0.1$, the density field is strongly unstably stratified at the beginning. The fluid in the bulk overturns with respect to the center height, which induces the sharp increase in $\Rey^{in}$. The bulk temperature quickly homogenizes due to faster diffusion, while the bulk salinity keeps in a stably stratified state for longer time period, as indicated by the high salinity near the bottom boundary and low salinity near the top boundary during $0<t<200$ in figure~\ref{fig:sthist1e10}c. As plumes grow from both plates and transport heat and salinity into the bulk, linear mean profiles are gradually established with upward gradients. The two different initial conditions lead to the same final states after $t>600$ with equal $\Lambda^{in}_b$ and $\Rey^{in}$, see figure~\ref{fig:timehist1e10}.

Above comparison indicates that the mixed initial condition generates the same final state as the linear initial condition providing that the final state is a single finger layer occupying the whole bulk. However, initial condition has non-trivial effects when staircases can develop in the system, as demonstrated in our previous work~\citep{ddcpnas2020}. One advantage of using mixed initial condition is that the transition time to the final state is shorter than that of the linear initial condition, which considerably saves the computational cost. It should be pointed out that the current findings are consistent with the 2D simulations in~\citet{ddcpnas2020}, where staircases were obtained in the range $\Ray_S>10^{12}$ at fixed $\Lambda=1.2$ for $\Pra=7$ and $\Sch=700$. Here staircases are observed at smaller $\Lambda$ for $\Ray_S=10^{12}$. Therefore, the critical value of $\Ray_S$ for the existence of staircase is smaller for lower $\Lambda$. Nevertheless, in this work we focus on the properties of single finger-layer state. The staircase sate will be left for future study.

\subsection{The fully developed state}

\begin{figure}
	\centerline{\includegraphics[width=\textwidth]{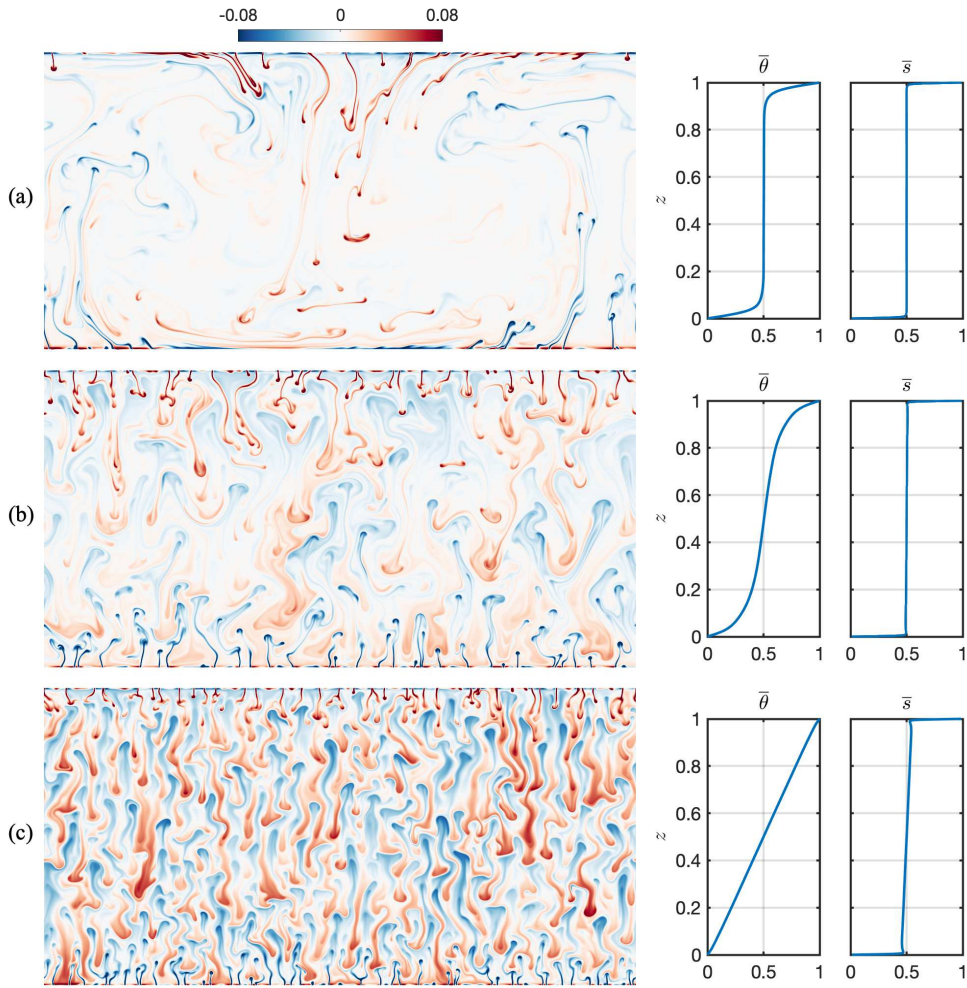}}
	\caption{The flow morphologies depicted by the contours of salinity anomaly for three cases with fixed $\Ray_S=10^{10}$ and (a) $\Lambda=0.01$, (b) $\Lambda = 0.1$, and (c) $\Lambda = 1$, respectively.}
	\label{fig:sa1e10}
\end{figure}

We now turn to the fully developed state of flow field. For every case we have run enough time to make the flow develop into the statistically steady state, and all the statistical data are calculated from this state over the time of $t_{stat}$ (see the Appendix). In particular, the case with $\Ray_S=10^8$ and $\Lambda=0.08$ has been run for over 10000 time units and the single finger-layer state persists. Previous experiments~\citep{kellner2014} and simulations~\citep{ddcpnas2016} both reveal that for fixed salinity difference between the two plates (i.e. fixed $\Ray_S$), as the temperature difference or equivalently the density ratio $\Lambda$ increases, the dominant flow structures change from large-scale convection rolls to slender salt fingers. The transition happens at about $\Lambda=0.03$. In figure~\ref{fig:sa1e10} we show those typical structures by the contours of salinity deviation from the mean salinity profile, i.e. $s'=s-\overline{s}(z)$. Hereafter, the overline stands for the temporal and horizontal average. Three cases with fixed $\Ray_S=10^{10}$ and increasing $\Lambda=0.01$, $0.1$, and $1.0$ are shown, respectively. For the field with $\Lambda=0.01$, a pair of large convection rolls emerge. In the horizontal locations apart from the ejection regions, the plumes developed from the near-wall boundary layers are tilted by the shear of convection motions. When $\Lambda$ increases to $0.1$, large convection rolls disappear and plumes develop into slender structures as they reach the bulk region. Near the boundary most of the plumes are vertically aligned. The slender structures in the bulk are distorted by the complex interactions among them. For $\Lambda=1$, the slender structures are well organized and vertically oriented salt fingers.

In figure~\ref{fig:sa1e10} we also plot the mean profiles for the two scalars, which are calculated by taking the temporal and horizontal average. For the case with the smallest $\Lambda=0.01$, see figure~\ref{fig:sa1e10}a, both scalars are perfectly mixed by the large scale flow motions and the mean temperature $\overline{\theta}$ and salinity $\overline{s}$ are independent of height, i.e. with very small vertical gradients. When the large-scale rolls are replaced by slender structures at $\Lambda=0.1$, the mean temperature profile $\overline{\theta}(z)$ exhibits notable gradient in the bulk, while the mean salinity profile $\overline{s}(z)$ also has a weak but non-zero gradient. For the case with $\Lambda=1$ shown in figure~\ref{fig:sa1e10}c, the mean temperature profile is very close to a linear one, and the non-zero gradient in salinity profile is clearly visible. 

\begin{figure}
	\centerline{\includegraphics[width=1\textwidth]{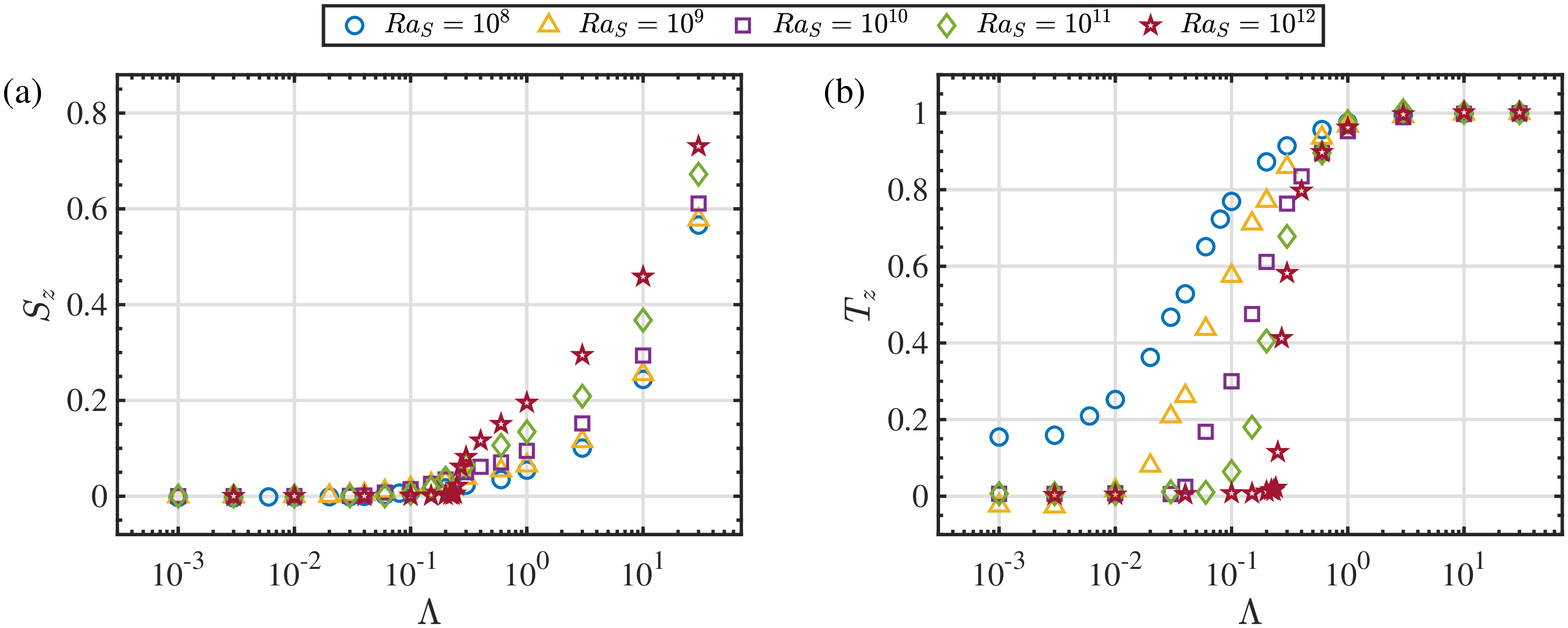}}
	\caption{The dependences of (a) the vertical gradient of mean salinity profile and (b) that of temperature profile in the bulk on the global density ratio $\Lambda$.}
	\label{fig:scalargradient}
\end{figure}

The vertical gradients of the mean temperature and salinity profiles are measured for all the cases and denoted by $T_z$ and $S_z$, respectively. Specifically, we calculate the slopes of $\overline{\theta}(z)$ and $\overline{s}(z)$ by linear fitting over the range of $0.25\le z \le 0.75$. Figure~\ref{fig:scalargradient} shows the dependences of the two slopes on the global density ratio $\Lambda$. For all five salinity Rayleigh numbers considered here, $S_z$ keeps almost zero when $\Lambda\le0.1$. For larger $\Lambda$, the slope $S_z$ increases with $\Lambda$. The bulk slope $T_z$ of the mean temperature profile starts to increase from zero at larger $\Lambda$ for higher $\Ray_S$, as shown in figure~\ref{fig:scalargradient}b. But $T_z$ reaches almost unity at $\Lambda\approx1$ for different $\Ray_S$, i.e. approaching a linear profile across the whole domain. Note that for $\Ray_S=10^8$, $T_z$ is not zero for the smallest $\Lambda$. For this small Rayleigh number, the convection rolls induced by the unstable salinity stratification are too weak to fully mix the thermal field and make it reach a homogeneous mean-temperature state.
\begin{figure}
	\centerline{\includegraphics[width=1\textwidth]{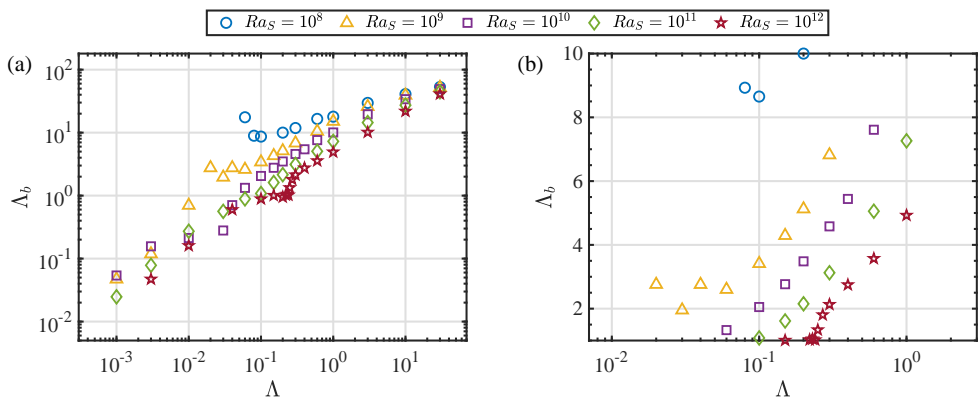}}
	\caption{The density ratio $\Lambda_b$ measured in the bulk versus the global density ratio. (a) the whole dataset except for the negative $\Lambda_b$ in the cases of $Ra_S=10^8$  and (b) zoom-in plot highlighting the cases with $\Lambda_b\in[1,10]$ which are the typical values of FDDC in the Ocean.}
	\label{fig:bulklambda}
\end{figure}

The bulk density ratio $\Lambda_b$ defined by the mean temperature and salinity gradients can be computed as $\Lambda_b=\Lambda T_z / S_z$, whose value is strongly affected by the dominant structures in the bulk. In figure~\ref{fig:bulklambda} we plot the variation of $\Lambda_b$ versus the global density ratio $\Lambda$. Generally  $\Lambda_b$ varies monotonically from around $0.025$ to $53$ with $\Lambda$ changing from $10^{-3}$ to $30$. Negative $\Lambda_b$ appears in the cases of $Ra_S=10^8$ for small $\Lambda$, which are not shown in the figure. Specifically,  $\Lambda_b$ changes from $1$ to $10$, which are the oceanic typical values of FDDC, only accompanied with a very little variation of $\Lambda$, as shown in figure \ref{fig:bulklambda}b. Based on $\Lambda_b$ and the mean salinity gradient $S_z$, we sort all cases into the convection type and the salt-finger type, which are shown in figure \ref{fig:params} by different symbols. In this study we identify a case as the salt-finger type if $\Lambda_b>1$ and $S_z>0.006$. The first criterion is a necessary condition for the salt-finger instability~\citep{stern1960}. The second criterion is also necessary here since for the convection-type of cases, both $T_z$ and $S_z$ in the bulk are very close to zero and their ratio can be artificially larger than unity. The criteria can also be quantitatively obtained by assuming that the horizontal wavelength of the salt fingers is smaller than $0.5H$, as we will shown in figure \ref{fig:xscale}a later. Note the threshold value of $\Lambda$ for the onset of salt-finger state varies slightly as $\Ray_S$ increases, changing from the lowest value of 0.04 to the highest value of 0.25. This variation was not revealed by our previous study for three-dimensional simulations~\citep{ddcpnas2016}, mainly due to the relatively low $\Ray_S$ and large step-size in $\Lambda$ used there.

\subsection{Characteristic lengths and different states of salt fingers}

\begin{figure}
	\centerline{\includegraphics[width=\textwidth]{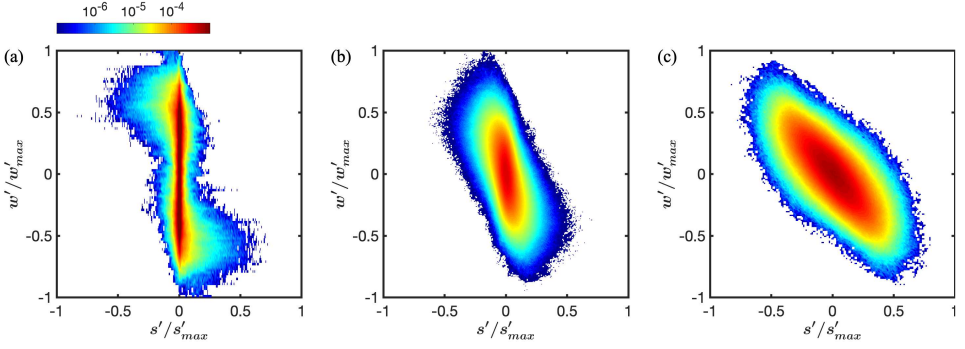}}
	\caption{Joint probability of the vertical velocity anomaly and the salinity anomaly normalized by their maximum values, respectively. The control parameters read $\Ray_S=10^{10}$ and (a) $\Lambda=0.01$, (b) $\Lambda = 0.1$, (c) $\Lambda = 1$.}
	\label{fig:pdf1e10}
\end{figure}

Based on the above criteria, the case shown in figure~\ref{fig:sa1e10}a is of the convection type and those in figures~\ref{fig:sa1e10}b and c are both of the salt-finger type. However, detailed investigations reveal that there is discrepancy between the flow fields in figures~\ref{fig:sa1e10}b and c. To demonstrate this, we calculate the joint probability density functions (pdfs) of $w'$ and $s'$ in the region $0.25\le z \le0.75$ for the three cases shown in figure~\ref{fig:sa1e10}. These joint pdfs are shown in figure~\ref{fig:pdf1e10}. When $\Lambda=0.01$, the pdf has a peak ridge along the axis $s'=0$ and over a wide range of $w'$. This region corresponds to low salinity anomaly with very different vertical velocity. There are also occasions with large positive (negative) salinity anomaly $s'$ associated with large negative (positive) vertical velocity $w'$, but the pdf is much lower. All these behaviours of pdf distribution in figure~\ref{fig:pdf1e10}a are consistent with the flow morphology of large convection rolls at $\Lambda=0.01$. The large convection rolls are mainly driven by the plumes growing from the boundary, instead of the local salinity anomaly in the bulk. On the contrary, when $\Lambda=1$ and the bulk is dominated by slender salt fingers as shown in figure~\ref{fig:sa1e10}c, the joint pdf is basically along the straight line of $w'/w'_{max}=-s'/s'_{max}$. The strong anti-correlation between $w'$ and $s'$ implies that the vertical velocity is very likely driven by the local salinity anomaly in the bulk which is carried by salt fingers.

For the case with $\Lambda=0.1$, figure~\ref{fig:sa1e10}b indicates that the flow structures in the bulk are more similar to salt fingers without large scale convection rolls. And the bulk density ratio $\Lambda_b$ is larger than unity. However, the joint pdf in figure~\ref{fig:pdf1e10}b exhibits a mixed nature of that for convection rolls and that for salt fingers. Specifically, the peak region of pdf is not along the axis $s'=0$, meanwhile the overall pattern is not along the anti-correlation line $w'/w'_{max}=-s'/s'_{max}$. Therefore, for the case with $\Lambda=0.1$, the bulk is in an intermediate state which is not entirely same as the salt finger state, even though the dominant flow structures are very similar to fingers.

\begin{figure}
	\centerline{\includegraphics[width=1\textwidth]{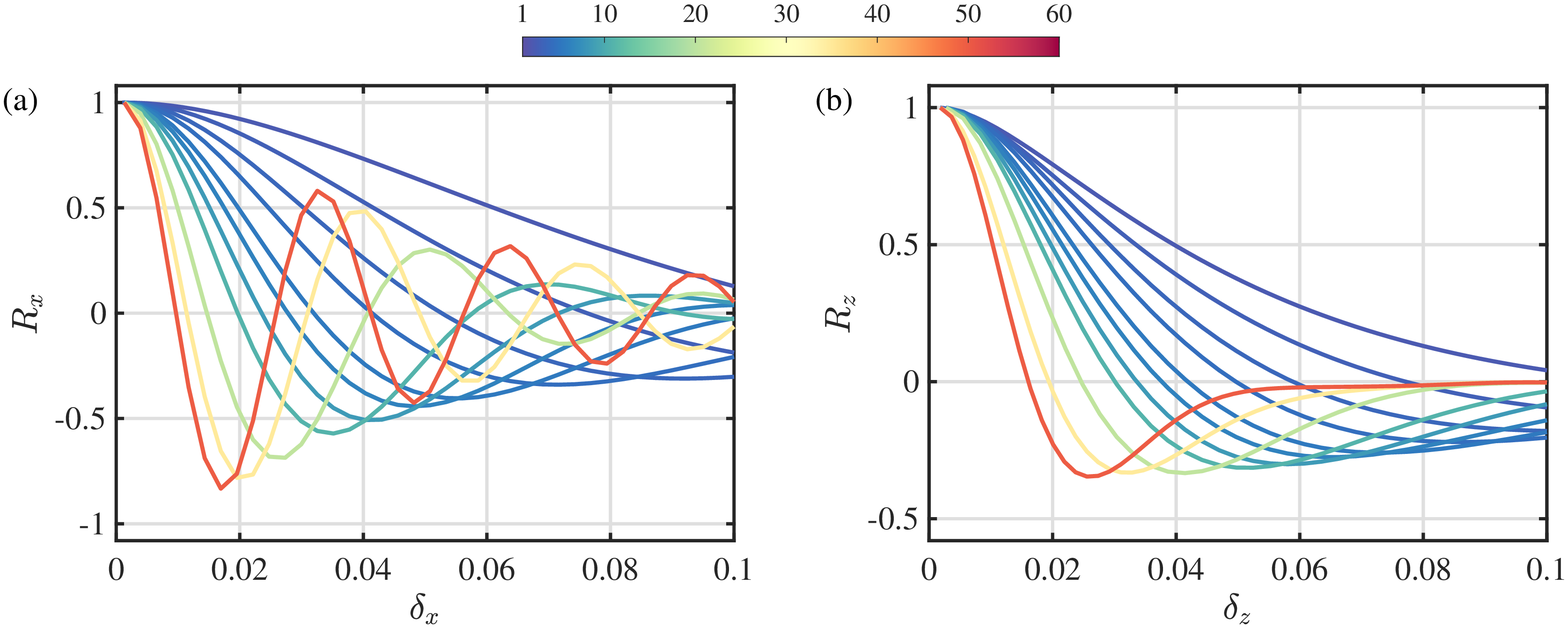}}
	\caption{(a) The horizontal auto-correlation functions $R_x$ versus the horizontal separation $\delta_x$  and (b) the vertical auto-correlation functions $R_z$ versus the vertical separation $\delta_z$ for the cases of salt-finger type with $Ra_S=10^{10}$. The colours are determined by the bulk density ratio $\Lambda_b$. }
	\label{fig:calscale}
\end{figure}

In order to distinguish the two states shown in figure~\ref{fig:pdf1e10}b and c, we investigate the characteristic length scales of the bulk structures for the cases of salt-finger type. The horizontal width and the vertical length can be extracted by using the auto-correlation functions of the vertical velocity $w$ which are defined as
\begin{equation}
R_x(\delta x) = \frac{\langle w(x,z,t) w(x+\delta x,z,t) \rangle_b } 
                     {\langle w^2(x,z,t) \rangle_b}, \quad
R_z(\delta z) = \frac{\langle w(x,z,t) w(x,z+\delta z,t) \rangle_b }
                     {\langle w^2(x,z,t) \rangle_b}.  
\end{equation}
Hereafter $\langle\,\rangle_b$ denotes the temporal and spatial average over the bulk region $0.25\le z \le 0.75$. Figure \ref{fig:calscale} demonstrates the behaviours of $R_x$ and $R_z$ for the salt-finger cases with $\Ray_S=10^{10}$ and $\Lambda_b>1$. For all the cases here, the auto-correlation curves always decrease to zero, and the horizontal intersect approximately equals to a quarter of the corresponding wavelength. The horizontal wavelength $\lambda_x$ and the vertical wavelength $\lambda_z$ are then calculated as four times the location of the first zero points of $R_x$ and $R_z$, respectively. 
\begin{figure}
	\centerline{\includegraphics[width=1\textwidth]{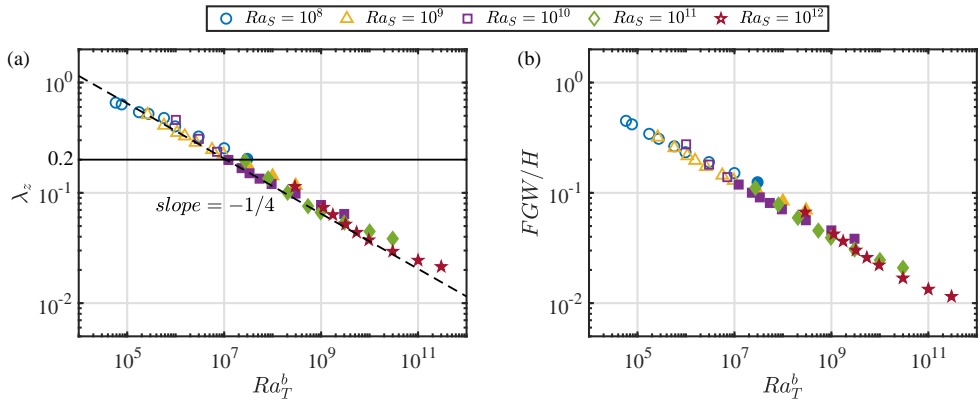}}
	\caption{(a) The vertical wavelength $\lambda_z$ versus the bulk temperature Rayleigh number $Ra_T^b$. (b) The ratio of FGW predicted by the linear model of~\citet{schmitt1979} to the domain height $H$. In (a) the horizontal solid line indicates the threshold value $\lambda_z=0.2$, and the dashed line denotes the $-1/4$ power-law scaling, respectively. The CSF cases are marked by open symbols, and FSF by closed ones, respectively.}
	\label{fig:zscale}
\end{figure}

We first look at the vertical wavelength $\lambda_z$. Figure~\ref{fig:zscale}a depicts the dependence of $\lambda_z$ on the bulk Rayleigh number of temperature $\Ray_T^b$ which is defined by the mean gradient of temperature profile in the bulk as $\Ray_T^b=\Ray_T T_z$. $\lambda_z$ decreases with $\Ray_T^b$ according to the scaling law $\lambda_z\sim\left(\Ray_T^b\right)^{-1/4}$. It is well known that the finger width is related to the mean thermal gradient as $d=(\kappa_T \nu/g\beta_T \partial_z\overline{ \theta })^{1/4}$~\citep{stern1960}, which corresponds to the same scaling law $d\sim\left(\Ray_T^b\right)^{-1/4}$. This implies that the ratio between the vertical wavelength $\lambda_z$ and finger width $d$ should be constant. Note that for small $\Ray_T^b$, $\lambda_z$ can be comparable to the domain height $H$. It can be expected that for these cases the boundary must affect the dynamics of salt fingers. Only those cases with $\lambda_z$ considerably smaller than $H$ have negligible influences on salt fingers in the bulk from the two boundaries. A practical threshold value for the current system is $\lambda_z=0.2$ as marked in figure~\ref{fig:zscale}a, which is equivalent to $\Ray_T^b\approx10^7$. With this threshold value we further divide the salt-finger regime into the confined salt-finger (CSF) regime with $\lambda_z>0.2$ and the free salt-finger (FSF) regime with $\lambda_z\le0.2$, respectively. These two regimes are illustrated by different types of symbol in the phase diagram shown in figure~\ref{fig:params}. For comparison, we also calculate the fastest growing wavelength (FGW) predicted by the linear model of~\citet{schmitt1979}, which is plotted in figure~\ref{fig:zscale}b. It can be seen that FGW is smaller than $\lambda_z$, and the threshold $\lambda_z=0.2$ corresponds to the FGW about $0.1H$.  
\begin{figure}
	\centerline{\includegraphics[width=1\textwidth]{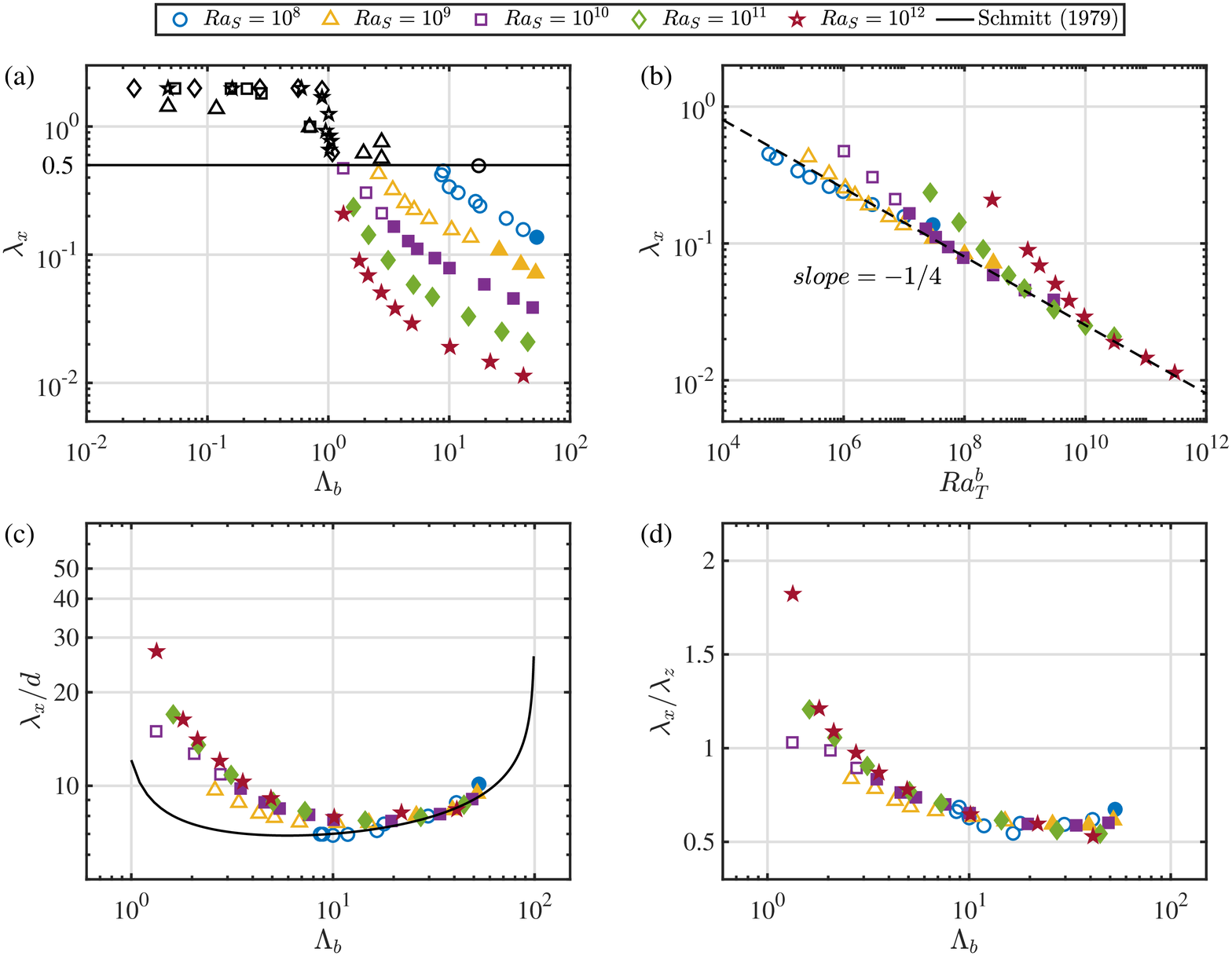}}
	\caption{The variation of the horizontal wavelength $\lambda_x$ for the cases of salt-finger type. (a) $\lambda_x$ versus the bulk density ratio $\Lambda_b$ with the horizontal solid line $\lambda_x=0.5$ which denotes the cases of convection type shown by black symbols. (b) $\lambda_x$ versus the bulk temperature Rayleigh number $Ra_T^b$ with the dashed line indicating the $-1/4$ power-law scaling. (c) $\lambda_x$ normalized by the finger scale $d$ versus $\Lambda_b$ with the solid line indicating the FGW predicted by the theoretical model of~\citet{schmitt1979}. (d) The aspect ratio $\lambda_x/\lambda_z$ versus $\Lambda_b$. The CSF cases are marked by open symbols, and FSF by closed ones, respectively.}
	\label{fig:xscale}
\end{figure}

Figure~\ref{fig:xscale} shows the behaviours of $\lambda_x$ for all the salt-finger cases with $\Lambda_b>1$ and $S_z>0.006$. The CSF and FSF cases are marked by open and solid symbols, respectively. In particular, in figure~\ref{fig:xscale}a the cases of convection type are also plotted by the black symbols. We can see that the salt-finger criteria of $\Lambda_b>1$ and $S_z>0.006$ guarantee that the horizontal wavelength is smaller than $0.5H$. Also after this threshold, $\lambda_x$ gradually decreases with $\Lambda_b$. Figure~\ref{fig:xscale}b indicates that for most cases $\lambda_x\sim\left(\Ray^b_T\right)^{-1/4}$ as suggested by the linear instability analysis.  Deviation from the power-law scaling can be observed for some cases, which usually have relatively small $\Lambda_b$. As $\Lambda_b$ approaches unity the salt-finger bulk is more turbulent and the nonlinear effects are stronger, which may cause the deviation. Figure~\ref{fig:xscale}c displays the ratio $\lambda_x/d$ versus $\Lambda_b$, which is compared with the theoretical prediction given by~\citet{schmitt1979}. $\lambda_x/d$ follows a single dependence on $\Lambda_b$ which agrees with the theoretical prediction for $\Lambda_b\ge10$. When $\Lambda_b<10$, the ratio is larger than the model prediction, which again can be attributed to the nonlinear effects at small bulk density ratio. The aspect ratio of salt fingers, measured by $\lambda_x/\lambda_z$, is plotted versus $\Lambda_b$ in figure~\ref{fig:xscale}d. As $\Lambda_b$ increases, the ratio gradually decreases and saturates. That is, the salt fingers shift from the blob-like shape at small $\Lambda_b$ to the slender shape at large $\Lambda_b$. When $\Lambda_b$ is large enough, the aspect ratio is nearly constant with $\lambda_z$ roughly twice the $\lambda_x$. Note the  previous asymptotic analysis indicates that the aspect ratio of salt fingers tends to unity when the density ratio approaches one~\citep{radko2008,vonHardenberg2010}, but in our wall-bounded model the aspect ratio exceeds unity for small bulk density ratio.

\subsection{The multi-layer state at $\Ray_S=10^{12}$}\label{sec:highra} 

Our previous work~\citep{ddcpnas2020} demonstrates that the staircase morphology exists for high Rayleigh numbers, and indeed it appears in some of the cases here with $\Ray_S=10^{12}$. Moreover, the mean salinity and temperature gradients in the bulk for $\Ray_S=10^{12}$ exhibit very abrupt transition as the global density ratio $\Lambda$ increases, see figure~\ref{fig:scalargradient}. Detailed investigations reveal that, during this transition the flow morphology undergoes interesting and complex changes which do not show up in other smaller $\Ray_S$ and will be discussed here in details.

\begin{figure}
	\centerline{\includegraphics[width=1\textwidth]{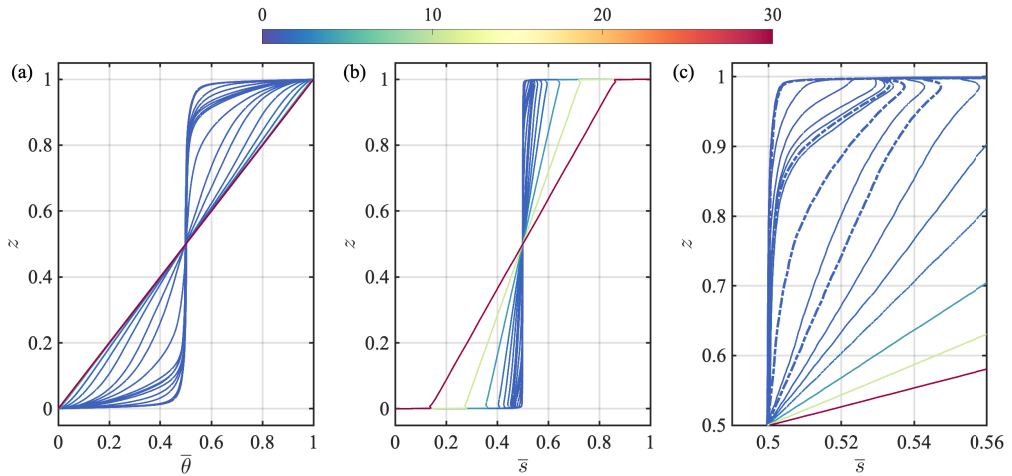}}
	\caption{The mean vertical profiles for (a) temperature and (b) salinity for all the cases with $\Ray_S=10^{12}$, coloured by the global density ratio $\Lambda$. Panel (c) shows the zoom-in plot of the upper part in (b) to amplify the development of the linear region next to the BL. The dash-dotted lines mark the four cases with $\Lambda=0.04, 0.23, 0.25$ and $0.3$ shown in figure~\ref{fig:w1e12}.}
    \label{fig:profile1e12}
\end{figure}

The changing of the mean scalar profiles for all the cases with $\Ray_S=10^{12}$ is shown in figure~\ref{fig:profile1e12}. As the global density ratio $\Lambda$ increases, the profiles shift from a convection state with homogeneous bulk to the salt-finger state with non-zero slope in the bulk. Since the diffusivity of temperature is much larger compared to that of salinity, the temperature profiles can reach the total conductive state for very large $\Lambda$. Closer inspection indicates that, during the transition from convection state to finger state, linear regions with non-zero slopes first develop next to the BLs, as highlighted in figure~\ref{fig:profile1e12}c. The two linear regions at upper and lower parts of the bulk gradually expand in height as the global density ratio $\Lambda$ increases. Eventually, the mean profiles in the whole bulk become linear with an almost uniform slope when $\Lambda$ is large enough.
\begin{figure}
	\centerline{\includegraphics[width=\textwidth]{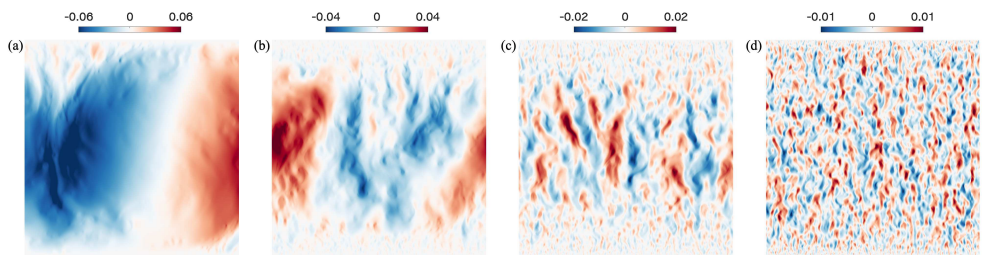}}
	\caption{The flow morphology depicted by the contours of vertical velocity for four cases with fixed $\Ray_S=10^{12}$ and (a) $\Lambda=0.04$, (b) $\Lambda = 0.23$, (c) $\Lambda = 0.25$, and (d) $\Lambda = 0.3$, respectively. In (a) only half the domain width is shown.}
	\label{fig:w1e12}
\end{figure}

These changes in profiles reflect the complex transition of the flow structures. In figure~\ref{fig:w1e12} we plot the contours of vertical velocity fields for four cases with different $\Lambda$ and different flow structures. For the case of $\Lambda=0.04$ the domain width is twice the widths of other three cases, and here only half the domain is shown. The corresponding mean profiles are marked in figure~\ref{fig:profile1e12}c by dash-dotted lines. For the case of $\Lambda=0.04$, the plumes which grow from BLs are quickly distorted by the shear associated with the large-scale convection rolls in the bulk. As $\Lambda$ increases to $0.23$, the plumes first extend vertically over a certain height and then enter the rolls. Now the central part of the bulk is still in the convection state but between the convection bulk and the BL there is a region filled with vertically aligned plumes or fingers, where the mean profile is linear. This kind of morphology resembles the multi-layer configuration of thermohaline staircase. For slightly higher $\Lambda=0.25$, the convection rolls are entirely replaced by the slender salt fingers and the bulk is already in the FSF state. Note that the scales of the bulk fingers are different from those of the plumes next to BLs. As $\Lambda$ increases even higher, the bulk fingers become thinner and taller, similar to those in the fully periodic domain. 

We search all the cases with $\Ray_S=10^{12}$ for those which have the multi-layer bulk as shown in figure \ref{fig:w1e12}b. Only three cases with $\Lambda=0.22, 0.23$ and $0.24$ exhibit this type of bulk morphology, and they are marked by stars in figure \ref{fig:params}. The exactly same flow state has been observed for large Rayleigh numbers in our previous study~\citep{ddcpnas2020}, where $\Lambda$ is fixed at $1.2$. For the 2D simulations there the multi-layer state exists at $\Ray_S=10^{12}$ for $\Sch=21$ but does not for $\Sch=700$. Here we observe this state for $\Sch=700$ and $\Ray_S=10^{12}$ at much smaller $\Lambda$. Therefore, starting from the mixed initial condition, the multi-layer state appears at larger $\Ray_S$ for larger $\Sch$ and fixed $\Lambda$, or at smaller $\Lambda$ for larger $\Sch$ and fixed $\Ray_S$ which is large enough. The existing conditions and dynamics of the staircase state are beyond the scope of current study, but definitely the subjects of future works.

\section{On the transport properties}\label{sec:trans} 

\begin{figure}
  \centerline{\includegraphics[width=\textwidth]{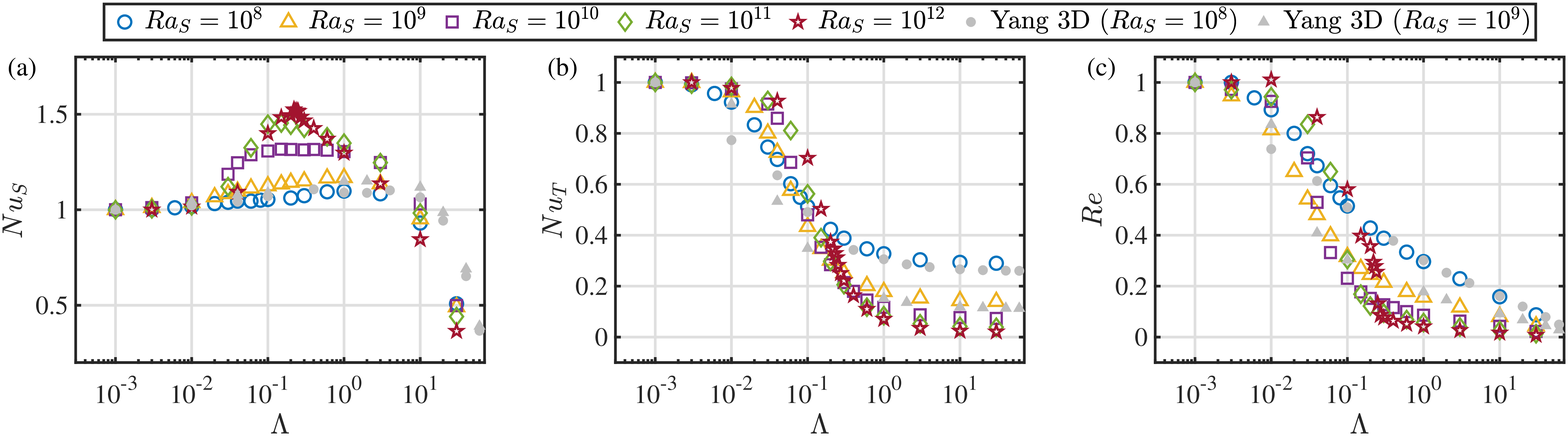}}
  \caption{The (a) salinity Nusselt number, (b) temperature Nusselt number, and (c) Reynolds number versus the density ratio, respectively. All quantities are normalized by the values of the case with smallest density ratio within each group. The grey symbols denote the results from 3D simulations of~\citet{ddcpnas2016}. }
  \label{fig:fluxratio}
\end{figure}

We now turn to the transport properties of the system. The key global responses include two Nusselt numbers and the Reynolds number as
\begin{equation}
  \Nus_S = \frac{\left|\overline{ w^* s^* } - \kappa_S \partial_z \overline{ s^* } \right|}{\kappa_S \Delta_S H^{-1}}, \quad
  \Nus_T = \frac{\left|\overline{ w^* \theta^* } - \kappa_T \partial_z \overline{ \theta^* } \right|}{\kappa_T \Delta_T H^{-1}}, \quad
  \Rey = \frac{U^*_{rms} H}{\nu}.  
\end{equation}
When the flow reaches the statistically steady state, $\Nus$ calculated by the above formula should be the same for arbitrary height, since under the horizontal periodic condition the net fluxes can only transport vertically. $U^*_{rms}$ is the root-mean-square (rms) value of the magnitude of velocity vector, which is computed over the entire domain. The dependences of these global responses on the global density ratio $\Lambda$ are displayed in figure~\ref{fig:fluxratio} for the five different $\Ray_S$. Note the quantities are normalized by the corresponding values of the smallest density ratio within each group.

The overall behaviours are very similar to those reported in our previous 3D simulations~\citep{ddcpnas2016}, as shown by the grey symbols in figure \ref{fig:fluxratio}. As $\Lambda$ increases, the salinity flux first increases and then quickly decreases. The enhancement of salinity flux is caused by the large-scale convection rolls  gradually being replaced by the well-organized salt fingers which can transport salinity more efficiently~\citep{kellner2014,ddcpnas2016}. Moreover, for higher $\Ray_S$ the salinity-flux enhancement is stronger. At $\Ray_S=10^{12}$ the increment of $\Nus_S$ can be as high as about $50\%$. The heat flux and the Reynolds number exhibit similar behaviours: They both first keep nearly constant and then quickly decrease towards very small values. Recall that the temperature gradient stabilizes the flow, then it is natural to expect that flow motions become weaker as $\Lambda$ increases. Both $\Nus_T$ and $\Rey$ decrease more abruptly for higher $\Ray_S$.

\begin{figure}
	\centerline{\includegraphics[width=1\textwidth]{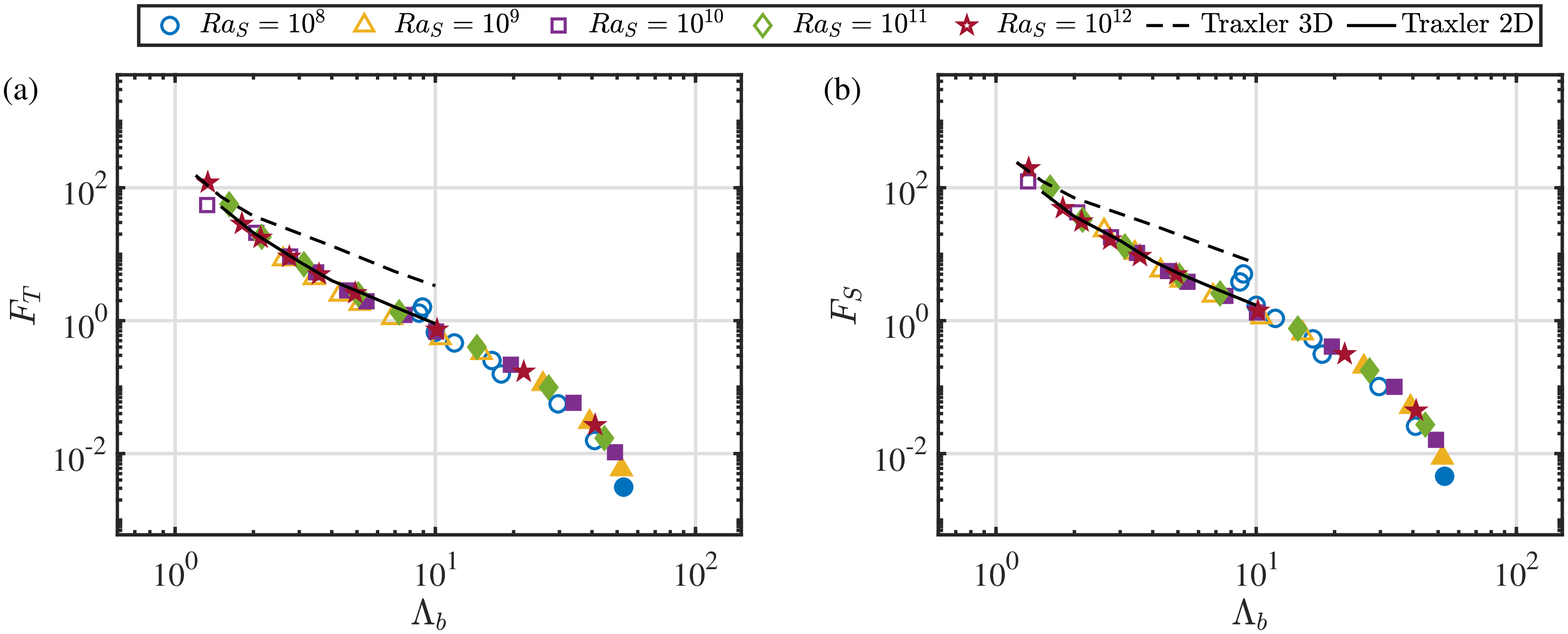}}
	\caption{The (a) heat flux and (b) salinity flux non-dimensionalized by the bulk scalar gradients versus the bulk density ratio, respectively. The dashed and solid lines denote the results from 3D and 2D periodic simulations of~\citet{traxler2011}. The CSF cases are marked by open symbols, and FSF by closed ones, respectively.}
	\label{fig:traxler1}
\end{figure}

To show the similarity between the results of the present wall-bounded model and those of the fully periodic model for the cases of salt-finger type, we redefine all the non-dimensional fluxes by the quantities measured within the bulk region, which are made as close as possible to the corresponding definitions in the periodic model. The results are then compared with those reported by~\citet{traxler2011} with the same fluid properties, namely $Pr=7$ and $Sc=700$. Firstly, the scalar fluxes non-dimensionalized by the bulk scalar gradients can be calculated as
\begin{equation}
F_T=\left| \frac{\langle w^*\theta^* \rangle_b }{\kappa_T \Delta_T H^{-1} T_z }\right|, \quad
F_S = \frac{1}{Le \Lambda_b}\left|\frac{\langle w^* s^*\rangle_b }{\kappa_S \Delta_S H^{-1}S_z}\right|.
\end{equation}
Here the bracket $\langle \rangle_b$ again denotes the temporal and spacial average in the bulk region $0.25\le z \le 0.75$. In~figure~\ref{fig:traxler1} we plot both heat and salinity fluxes versus the bulk density ratio. The 2D and 3D results of~\citet{traxler2011} are also included for direct comparison. Near-perfect agreement between two studies of 2D simulations is obtained over the common range of $\Lambda_b$ for both scalar fluxes, although with a little deviation for CSF. That is, when the salt fingers emerge in the wall-bounded flow and if all the quantities are expressed in the measured bulk values, the same dependence of fluxes on density ratio applies to both wall-bounded model and fully periodic model. Figure~\ref{fig:traxler1} also demonstrates that, as $\Lambda_b$ increases from $1$ to about $60$, both the non-dimensional heat and salinity fluxes decrease from above $10^2$ to below $10^{-2}$, indicating that the main transport mode shifts from the turbulent convection to the molecular diffusion. 

\begin{figure}
	\centerline{\includegraphics[width=\textwidth]{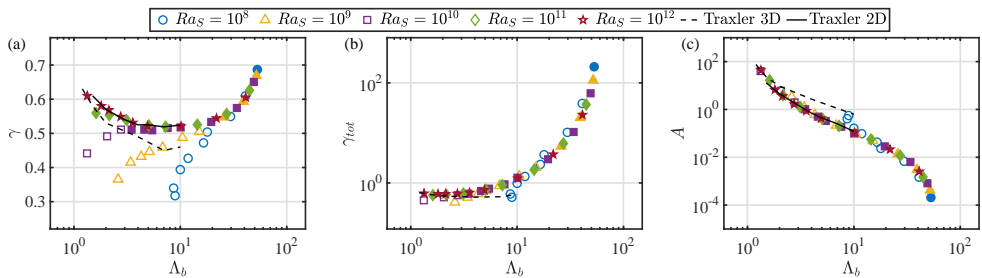}}
	\caption{(a) The flux ratio $\gamma$, (b) the total flux ratio $\gamma_{tot}$ and (c) the Stern number $A$ versus the bulk density ratio $\Lambda_b$. The dashed and solid lines denote the results from 3D and 2D periodic simulations of~\citet{traxler2011}. The CSF cases are marked by open symbols, and FSF by closed ones, respectively.}
	\label{fig:traxler2}
\end{figure}

We further examine the turbulent flux ratio, the total flux ratio and the Stern number measured from the bulk as, respectively,
\begin{equation}
  \gamma= \frac{F_T}{F_S}, \quad
  \gamma_{tot}= \frac{\beta_T \langle w^*\theta^*-\kappa_T \partial_z\theta^* \rangle_b}
                     {\beta_S \langle w^*s^*-\kappa_S \partial_z s^* \rangle_b}, \quad
  A=\frac{F_S-F_T}{Pr(1/\Lambda_b-1)}.
\end{equation}
The turbulent flux ratio represents the ratio of density flux caused by the convective heat transfer to that by the convective salt transfer, while the total flux ratio includes the diffusion part. The Stern number controls the collective instability of salt-finger layers, which is a large-scale secondary instability related to the gravity waves~\citep{stern1969,stern2001}. The dependence of $\gamma$ on $\Lambda_b$ for all the cases of salt-finger type is shown in figure~\ref{fig:traxler2}a. For the FSF type, $\gamma$ first decreases and then increases as $\Lambda_b$ becomes smaller and approaches unity. $\gamma$ reaches the minimum at around $\Lambda_b=10$. This variation is consistent with the 2D periodic simulations~\citep{traxler2011}, see the comparison between the solid symbols and the solid line. For the CSF type, although the two fluxes $F_S$ and $F_T$ are very close to those in the periodic simulations shown in figure \ref{fig:traxler1}, the variation of $\gamma$ deviates from the trend of FSF type as $\Lambda_b$ decreases. The deviation starts at higher $\Lambda_b$ for smaller $\Ray_S$. This deviation is attributed to the fact that salt fingers are influenced by the energetic boundary plumes in CSF state, slightly changing $F_T$ and $F_S$ but making big effects on their ratio. Figure \ref{fig:traxler2}b shows the variation of $\gamma_{tot}$ with $\Lambda_b$. For FSF type, $\gamma_{tot}$ also converges into one curve for different $Ra_S$ and increases monotonically with $\Lambda_b$. For small $\Lambda_b$ the data is close to those of the 3D periodic simulations~\citep{traxler2011} (the 2D data of $\gamma_{tot}$ is not given in that paper), but lack of a range with decreasing $\gamma_{tot}$. Small deviations still exist for CSF type. The dependence of the Stern number on $\Lambda_b$ shows the same trend with the scalar fluxes, and again quantitatively agreement with the periodic simulations~\citep{traxler2011}, see figure \ref{fig:traxler2}c. 

Above discussions confirm that, once the wall-bounded FDDC is in the FSF regime, the transport behaviours of salt-fingers are exactly the same as those in periodic domain. However, the solid boundary in wall-bounded model can alter the properties related to the large-scale secondary instability. For instance, the $\gamma$ instability can happen in the periodic model when the total flux ratio decreases with the density ratio~\citep{traxler2011}. Here in the 2D wall-bounded FDDC in the FSF regime, such condition is not satisfied for the parameters explored. The collective instability, which should develop when $A$ exceeds unity according to the linear theory~\citep{stern1969}, is also not observed in the current simulations. Note the largest Stern number obtained for the FSF cases is about 100. The inclined gravity waves in periodic FDDC reported by~\citet{stellmach2011} are absent in all the cases here, which is expectable since the solid boundaries prevents the long-time propagation of such inclined wave. Another reason for the absence of the collective instability may be the fact that the domain height $H$ is too small for those cases with large Stern number (see figure \ref{fig:zscale}b). As in periodic simulations~\citep{traxler2011, stellmach2011}, $H$ needs to be larger than about 25 FGW to produce the large-scale secondary instability.

\section{Conclusions}\label{sec:conclusion} 

In summary, we conduct systematic 2D DNS of the wall-bounded FDDC for the same fluid properties of seawater. For several fixed salinity Rayleigh numbers in the range of $10^8\le\Ray_S\le10^{12}$, we gradually increase the density ratio $\Lambda$, or equivalently the temperature Rayleigh number. The changing in the flow morphology and transport properties are investigated. Especially, we establish the correspondence between the wall-bounded model and periodic model of FDDC.

At small $\Lambda$, the flow is similar to RB flow with the bulk dominated by the large-scale convection rolls. As $\Lambda$ increases, the large-scale convection rolls are replaced by the slender salt-fingers in the bulk and the flow enters the SF regime. Meanwhile, the bulk density ratio $\Lambda_b$, which is defined by the mean gradients of temperature and salinity at the middle part of domain, exceeds unity. These observations are consistent with previous experiments and simulations~\citep{hage2010,kellner2014,ddcpnas2016}. However, detailed analyses reveal that the SF regime can be further divided into the CSF and FSF regimes with different salt-finger states in the bulk. In the CSF regime the characteristic height of salt fingers is comparable with the domain height. While in the FSF regime the characteristic height is much smaller than the domain height and the vertical motions in the bulk are mainly driven by the local salinity anomaly. Therefore, for the FSF regime the effects of solid boundary on salt fingers in the bulk is much weaker. We measure the characteristic height by the vertical wavelength $\lambda_z^*$ which is calculated by using the auto-correlation function of vertical velocity, and propose the criterion $\lambda_z^*<0.2H$ for the FSF regime. This criterion corresponds to $\Ray_T^b>10^7$, where the bulk Rayleigh number of temperature $\Ray_T^b$ is defined with the mean temperature gradient in the bulk and the domain height.

The salinity flux first increases and then decreases as the density ratio becomes larger, meanwhile both the heat flux and flow velocity decrease monotonically. The salinity-flux enhancement at the intermediate density ratios becomes stronger for larger $\Ray_S$. For the cases in the FSF regime, if the salinity and heat fluxes are non-dimensionalized by the respective mean conductive fluxes in the bulk, their dependences on the bulk density ratio quantitatively agree with those obtained in periodic FDDC, e.g. those reported in~\citet{traxler2011}. The same observation is obtained for the density flux ratio, the total density flux ratio and the Stern number which controls the collective instability. Therefore, for the wall-bounded model in the FSF regime, the transport properties are the same as those in periodic domain if all quantities are expressed by the bulk values.

In all the cases within the FSF regime we do not observe the development of the secondary instabilities, such as the $\gamma$ instability and collective instability. For the FSF regime there is not a range in which the total density flux ratio decreases with the bulk density ratio, which is the requirement for the unset of $\gamma$ instability. Also the horizontal boundaries at the top and bottom prevent the long-time propagation of inclined gravity-wave modes. The 2D nature and the limited domain heights of the simulations may also affect the properties of secondary instability. And higher Rayleigh numbers may be needed for these large-scale instabilities to occur.

The single salt-finger layers in the current simulations do not spontaneously break into the multi-layer staircase state. However, starting from the mixed initial condition, we do obtain the multi-layer state in a very limited range of $\Lambda$ for the highest salinity Rayleigh number $\Ray_S=10^{12}$. Searching for the parameter range and identifying the condition where the staircase state exists are crucial to fully understand the physical mechanism and the evolution dynamics of fingering thermohaline staircases  in the Ocean. The wall-bounded model provides a useful system for these purposes, which are the subjects of our undergoing work.

\appendix

\section{Summary of Numerical Details} 

In the following tables we summarize the numerical details and key responses for all the simulations. Each table corresponds to one salinity Rayleigh number. Columns from left to right are the global density ratio $\Lambda$ defined by the temperature and salinity differences between the two plates, the aspect ratio $\Gamma$ of the domain, the resolution of the base mesh $(N_x,~ N_z)$, the refinement factors $(m_x,~m_z)$ of the refined mesh, the two Nusselt numbers $Nu_S$ and $Nu_T$, the Reynolds numbers defined by the rms of total velocity $Re$, by the rms of $x$-velocity $Re_x$, and by the rms of $z$-velocity $Re_z$, the density ratio measured at the bulk $\Lambda_b$ of the fully developed flow, the vertical gradients of the mean salinity and temperature, the statistical time in the fully developed state, respectively. For all cases the fluid properties of seawater are used, namely $Pr=7$ and $Sc=700$.
\begin{table}
\begin{center}
\def~{\hphantom{0}}
\begin{tabular}{ccccccccccccc}
 $\Lambda$  & $\Gamma$ & $N_x(m_x)$ & $N_z(m_z)$ & $Nu_S$ & $Nu_T$ &
 $Re$ & $Re_z$ & $Re_x$ & $\Lambda_b$ & $\overline{S}_z$ & $\overline{T}_z$ & $t_{stat}$\\[3pt]
  0.001  &  8.0  &   720(4)  &   192(2)  &    28.14  &    3.459  &    5.593  &    4.583 &   3.200  &  -0.0565  &  -0.003  &  0.154  &  1000 \\ 
0.003  &  8.0  &   720(4)  &   192(2)  &    28.20  &    3.426  &    5.594  &    4.461 &   3.370  &   -0.135  &  -0.004  &  0.159  &  1000 \\ 
0.006  &  8.0  &   720(4)  &   192(2)  &    28.43  &    3.309  &    5.254  &    4.289 &   3.029  &   -0.858  &  -0.001  &  0.209  &  1000 \\ 
0.01  &  8.0  &   720(4)  &   192(2)  &    28.62  &    3.191  &    4.992  &    4.113 &   2.824  &    -1.29  &  -0.002  &  0.252  &  1000 \\ 
0.02  &  8.0  &   720(4)  &   192(2)  &    29.05  &    2.883  &    4.474  &    3.713 &   2.492  &    -6.36  &  -0.001  &  0.362  &  1000 \\ 
0.03  &  8.0  &   720(4)  &   192(2)  &    29.24  &    2.581  &    4.029  &    3.346 &   2.241  &    -11.0  &  -0.001  &  0.467  &  1000 \\ 
0.04  &  8.0  &   720(4)  &   192(2)  &    29.42  &    2.413  &    3.763  &    3.136 &   2.079  &    -49.1  &  -0.000  &  0.528  &  1000 \\ 
0.06  &  8.0  &   720(4)  &   192(2)  &    29.41  &    2.083  &    3.326  &    2.795 &   1.801  &     17.5  &  0.002  &  0.651  &  1000 \\ 
0.08  &  8.0  &   720(4)  &   192(2)  &    29.55  &    1.898  &    3.058  &    2.584 &   1.635  &     8.93  &  0.006  &  0.723  &  10000 \\ 
0.1  &  8.0  &   720(4)  &   192(2)  &    29.66  &    1.773  &    2.879  &    2.437 &   1.530  &     8.65  &  0.009  &  0.769  &  1000 \\ 
0.2  &  8.0  &   720(4)  &   192(2)  &    29.87  &    1.464  &    2.396  &    2.061 &   1.220  &     10.0  &  0.017  &  0.873  &  1000 \\ 
0.3  &  8.0  &   720(4)  &   192(2)  &    30.21  &    1.344  &    2.172  &    1.886 &   1.077  &     11.8  &  0.023  &  0.914  &  1000 \\ 
0.6  &  8.0  &   720(4)  &   192(2)  &    30.79  &    1.199  &    1.866  &    1.634 &  0.9009  &     16.5  &  0.035  &  0.957  &  1000 \\ 
1  &  8.0  &   720(4)  &   240(2)  &    30.89  &    1.132  &    1.658  &    1.443 &  0.8156  &     17.9  &  0.054  &  0.976  &  1000 \\ 
3  &  8.0  &   720(4)  &   240(2)  &    30.49  &    1.050  &    1.282  &    1.124 &  0.6156  &     29.6  &  0.101  &  0.993  &  2000 \\ 
10  &  8.0  &   720(4)  &   240(2)  &    26.18  &    1.015  &   0.8859  &   0.7738 &  0.4309  &     41.0  &  0.244  &  0.999  &  4000 \\ 
30  &  8.0  &   720(4)  &   240(2)  &    14.29  &    1.003  &   0.4893  &   0.4165 &  0.2567  &     53.0  &  0.566  &  1.000  &  5000 \\ 
\end{tabular}
\caption{Numerical details and key responses for the group of cases with $Ra_S=10^8$.}
\label{tab:ras08}
\end{center}
\end{table}

\begin{table}
\begin{center}
\def~{\hphantom{0}}
\begin{tabular}{ccccccccccccc}
 $\Lambda$  & $\Gamma$ & $N_x(m_x)$ & $N_z(m_z)$ & $Nu_S$ & $Nu_T$ &
 $Re$ & $Re_z$ & $Re_x$ & $\Lambda_b$ & $\overline{S}_z$ & $\overline{T}_z$ & $t_{stat}$\\[3pt]
  0.001  &  5.0  &   768(4)  &   288(2)  &    53.91  &    7.210  &    27.77  &    20.77 &   18.41  &   0.0471  &  -0.001  &  -0.024  &   500 \\ 
0.003  &  5.0  &   768(4)  &   288(2)  &    54.51  &    7.203  &    26.29  &    19.25 &   17.89  &    0.118  &  -0.001  &  -0.027  &   500 \\ 
0.01  &  5.0  &   768(4)  &   288(2)  &    55.90  &    6.940  &    22.59  &    17.10 &   14.72  &    0.700  &  0.000  &  0.015  &   600 \\ 
0.02  &  5.0  &   768(4)  &   288(2)  &    57.54  &    6.499  &    18.04  &    14.50 &   10.70  &     2.76  &  0.001  &  0.079  &   800 \\ 
0.03  &  5.0  &   768(4)  &   288(2)  &    58.43  &    5.772  &    15.04  &    12.38 &   8.527  &     1.96  &  0.003  &  0.208  &   600 \\ 
0.04  &  5.0  &   768(4)  &   288(2)  &    59.13  &    5.228  &    13.34  &    10.93 &   7.646  &     2.76  &  0.004  &  0.261  &   800 \\ 
0.06  &  5.0  &   768(4)  &   288(2)  &    60.01  &    4.151  &    11.03  &    9.035 &   6.315  &     2.60  &  0.010  &  0.437  &  1000 \\ 
0.1  &  5.0  &   768(4)  &   288(2)  &    60.63  &    3.128  &    8.768  &    7.214 &   4.980  &     3.41  &  0.017  &  0.575  &   800 \\ 
0.15  &  5.0  &   768(4)  &   288(2)  &    61.16  &    2.501  &    7.469  &    6.209 &   4.149  &     4.29  &  0.025  &  0.712  &  1000 \\ 
0.2  &  5.0  &   768(4)  &   288(2)  &    61.64  &    2.172  &    6.817  &    5.706 &   3.729  &     5.12  &  0.030  &  0.771  &  1000 \\ 
0.3  &  5.0  &   768(4)  &   288(2)  &    62.02  &    1.821  &    5.956  &    5.053 &   3.151  &     6.82  &  0.038  &  0.859  &   800 \\ 
0.6  &  5.0  &   768(4)  &   288(2)  &    62.84  &    1.454  &    4.955  &    4.266 &   2.520  &     10.5  &  0.054  &  0.936  &  1000 \\ 
1  &  5.0  &   768(4)  &   288(2)  &    62.96  &    1.286  &    4.336  &    3.784 &   2.118  &     15.1  &  0.064  &  0.967  &  1200 \\ 
3  &  5.0  &   768(4)  &   288(2)  &    61.21  &    1.103  &    3.262  &    2.883 &   1.525  &     25.9  &  0.115  &  0.992  &  2400 \\ 
10  &  4.0  &   768(4)  &   384(1)  &    51.31  &    1.029  &    2.227  &    1.979 &   1.021  &     39.2  &  0.255  &  0.998  &  4000 \\ 
30  &  4.0  &   768(4)  &   384(1)  &    26.50  &    1.006  &    1.211  &    1.061 &  0.5832  &     51.9  &  0.578  &  1.000  &  4000 \\ 
\end{tabular}
\caption{Numerical details and key responses for the group of cases with $Ra_S=10^9$.}
\label{tab:ras09}
\end{center}
\end{table}

\begin{table}
\begin{center}
\def~{\hphantom{0}}
\begin{tabular}{ccccccccccccc}
 $\Lambda$  & $\Gamma$ & $N_x(m_x)$ & $N_z(m_z)$ & $Nu_S$ & $Nu_T$ &
 $Re$ & $Re_z$ & $Re_x$ & $\Lambda_b$ & $\overline{S}_z$ & $\overline{T}_z$ & $t_{stat}$\\[3pt]
  0.001  &  2.0  &  1024(4)  &   768(3)  &    97.74  &    13.77  &    131.5  &    91.62 &   94.29  &   0.0539  &  0.000  &  0.006  &   600 \\ 
0.003  &  2.0  &  1024(4)  &   768(3)  &    98.85  &    13.75  &    127.9  &    89.26 &   91.65  &    0.157  &  0.000  &  0.006  &   600 \\ 
0.01  &  2.0  &  1024(4)  &   768(3)  &    101.3  &    13.42  &    121.7  &    84.69 &   87.43  &    0.212  &  0.000  &  0.007  &   600 \\ 
0.03  &  2.0  &  1024(4)  &   768(3)  &    115.9  &    12.60  &    92.54  &    65.05 &   65.74  &    0.279  &  0.001  &  0.005  &   800 \\ 
0.04  &  2.0  &  1024(4)  &   768(3)  &    121.8  &    11.83  &    69.57  &    52.88 &   45.04  &    0.706  &  0.001  &  0.024  &   800 \\ 
0.06  &  2.0  &  1024(4)  &   768(3)  &    125.9  &    9.449  &    43.64  &    34.80 &   26.23  &     1.33  &  0.008  &  0.168  &   800 \\ 
0.1  &  2.0  &  1024(4)  &   768(3)  &    127.8  &    6.609  &    30.38  &    24.65 &   17.73  &     2.05  &  0.015  &  0.300  &  2800 \\ 
0.15  &  2.0  &   768(4)  &   512(3)  &    128.7  &    4.856  &    23.33  &    18.94 &   13.60  &     2.77  &  0.026  &  0.475  &  1000 \\ 
0.2  &  2.0  &   768(4)  &   512(3)  &    128.6  &    3.914  &    19.98  &    16.24 &   11.63  &     3.48  &  0.035  &  0.612  &  1000 \\ 
0.3  &  2.0  &   768(4)  &   512(3)  &    128.4  &    2.946  &    16.72  &    13.76 &   9.490  &     4.58  &  0.050  &  0.764  &  1000 \\ 
0.4  &  2.0  &   768(4)  &   512(3)  &    128.6  &    2.475  &    15.23  &    12.57 &   8.585  &     5.44  &  0.061  &  0.835  &  1000 \\ 
0.6  &  2.0  &   768(4)  &   512(3)  &    128.2  &    2.002  &    13.17  &    11.06 &   7.136  &     7.61  &  0.071  &  0.896  &  1000 \\ 
1  &  2.0  &   768(4)  &   512(3)  &    127.5  &    1.608  &    11.29  &    9.652 &   5.848  &     10.1  &  0.095  &  0.953  &  1600 \\ 
3  &  2.0  &   768(4)  &   512(2)  &    121.9  &    1.206  &    8.259  &    7.285 &   3.889  &     19.5  &  0.152  &  0.989  &  2400 \\ 
10  &  2.0  &   768(4)  &   768(1)  &    100.7  &    1.056  &    5.592  &    5.025 &   2.452  &     34.0  &  0.294  &  0.998  &  3200 \\ 
30  &  2.0  &   768(4)  &   768(1)  &    48.72  &    1.010  &    2.925  &    2.616 &   1.306  &     49.1  &  0.611  &  1.000  &  5000 \\ 
\end{tabular}
\caption{Numerical details and key responses for the group of cases with $Ra_S=10^{10}$.}
\label{tab:ras10}
\end{center}
\end{table}

\begin{table}
\begin{center}
\def~{\hphantom{0}}
\begin{tabular}{ccccccccccccc}
 $\Lambda$  & $\Gamma$ & $N_x(m_x)$ & $N_z(m_z)$ & $Nu_S$ & $Nu_T$ &
 $Re$ & $Re_z$ & $Re_x$ & $\Lambda_b$ & $\overline{S}_z$ & $\overline{T}_z$ & $t_{stat}$\\[3pt]
  0.001  &  2.0  &  3072(4)  &  1280(3)  &    186.1  &    26.31  &    516.7  &    359.5 &   371.0  &   0.0247  &  0.000  &  0.007  &   500 \\ 
0.003  &  2.0  &  3072(4)  &  1280(3)  &    187.0  &    26.13  &    501.8  &    349.4 &   360.2  &   0.0782  &  0.000  &  0.006  &   500 \\ 
0.01  &  2.0  &  2560(4)  &  1280(3)  &    189.9  &    25.90  &    487.6  &    339.5 &   349.9  &    0.272  &  0.000  &  0.007  &   600 \\ 
0.03  &  2.0  &  2560(4)  &  1280(3)  &    208.6  &    24.45  &    431.7  &    300.2 &   310.3  &    0.562  &  0.001  &  0.011  &   600 \\ 
0.06  &  2.0  &  2560(4)  &  1280(3)  &    246.5  &    21.34  &    335.8  &    234.2 &   240.7  &    0.887  &  0.001  &  0.010  &   800 \\ 
0.1  &  1.0  &  1024(4)  &  1024(3)  &    269.5  &    14.80  &    157.5  &    116.6 &   105.1  &     1.08  &  0.006  &  0.064  &  1000 \\ 
0.15  &  1.0  &  1024(4)  &  1024(3)  &    270.6  &    10.27  &    87.11  &    70.00 &   51.77  &     1.61  &  0.017  &  0.180  &  1000 \\ 
0.2  &  1.0  &  1024(4)  &  1024(3)  &    268.5  &    7.828  &    63.06  &    50.64 &   37.55  &     2.15  &  0.038  &  0.405  &  1000 \\ 
0.3  &  1.0  &  1024(4)  &  1024(3)  &    265.5  &    5.418  &    48.59  &    39.02 &   28.95  &     3.13  &  0.065  &  0.678  &  1000 \\ 
0.6  &  1.0  &  1024(4)  &  1024(3)  &    257.2  &    3.092  &    35.55  &    29.05 &   20.49  &     5.06  &  0.106  &  0.896  &  1000 \\ 
1  &  1.0  &   768(4)  &   864(3)  &    251.0  &    2.214  &    29.46  &    24.52 &   16.33  &     7.26  &  0.135  &  0.978  &   800 \\ 
3  &  1.0  &   768(4)  &   864(3)  &    232.0  &    1.377  &    20.51  &    17.81 &   10.17  &     14.4  &  0.209  &  1.006  &  1200 \\ 
10  &  1.0  &   768(4)  &   864(2)  &    182.9  &    1.097  &    13.34  &    11.92 &   5.989  &     27.3  &  0.367  &  1.002  &  2000 \\ 
30  &  1.0  &   768(4)  &   864(2)  &    82.04  &    1.016  &    6.808  &    6.139 &   2.943  &     44.6  &  0.672  &  1.001  &  4000 \\ 
\end{tabular}
\caption{Numerical details and key responses for the group of cases with $Ra_S=10^{11}$.}
\label{tab:ras11}
\end{center}
\end{table}

\begin{table}
\begin{center}
\def~{\hphantom{0}}
\begin{tabular}{ccccccccccccc}
 $\Lambda$  & $\Gamma$ & $N_x(m_x)$ & $N_z(m_z)$ & $Nu_S$ & $Nu_T$ &
 $Re$ & $Re_z$ & $Re_x$ & $\Lambda_b$ & $\overline{S}_z$ & $\overline{T}_z$ & $t_{stat}$\\[3pt]
  0.003  &  2.0  &  3072(8)  &  3072(3)  &    370.5  &    48.36  &    1883.  &    1307. &   1355.  &   0.0472  &  0.000  &  0.003  &   200 \\ 
0.01  &  2.0  &  3072(8)  &  3072(3)  &    376.3  &    47.30  &    1903.  &    1308. &   1381.  &    0.160  &  0.000  &  0.004  &   200 \\ 
0.04  &  2.0  &  3072(8)  &  3072(3)  &    405.0  &    44.84  &    1626.  &    1131. &   1167.  &    0.598  &  0.000  &  0.005  &   240 \\ 
0.1  &  2.0  &  3072(8)  &  3072(3)  &    518.6  &    34.00  &    1092.  &    762.3 &   781.7  &    0.886  &  0.001  &  0.008  &   400 \\ 
0.15  &  2.0  &  3072(8)  &  3072(3)  &    549.7  &    24.32  &    748.7  &    514.7 &   541.8  &     1.01  &  0.001  &  0.007  &   400 \\ 
0.2  &  1.0  &  2560(4)  &  3072(3)  &    553.6  &    18.08  &    671.3  &    473.5 &   466.1  &    0.950  &  0.003  &  0.012  &   800 \\ 
0.22  &  1.0  &  2560(4)  &  2560(3)  &    564.9  &    16.78  &    554.9  &    401.2 &   379.0  &     1.02  &  0.003  &  0.016  &   800 \\ 
0.23  &  1.0  &  2560(4)  &  2560(3)  &    561.9  &    15.92  &    523.5  &    389.3 &   348.5  &     1.05  &  0.003  &  0.014  &   800 \\ 
0.24  &  1.0  &  2560(4)  &  2560(3)  &    562.6  &    15.10  &    481.5  &    340.2 &   338.5  &     1.02  &  0.005  &  0.022  &   800 \\ 
0.25  &  1.0  &  2048(4)  &  2048(3)  &    561.5  &    13.67  &    247.5  &    185.2 &   163.8  &     1.33  &  0.022  &  0.115  &  1000 \\ 
0.27  &  1.0  &  2048(4)  &  2048(3)  &    550.3  &    12.12  &    163.0  &    129.2 &   99.44  &     1.81  &  0.062  &  0.412  &  1000 \\ 
0.3  &  1.0  &  2048(4)  &  2048(3)  &    543.4  &    10.72  &    143.8  &    114.6 &   86.91  &     2.13  &  0.082  &  0.582  &  1000 \\ 
0.4  &  1.0  &  2048(4)  &  2048(3)  &    528.5  &    7.918  &    119.6  &    95.33 &   72.15  &     2.75  &  0.116  &  0.797  &  1000 \\ 
0.6  &  1.0  &  2048(4)  &  2048(3)  &    507.8  &    5.325  &    97.33  &    77.65 &   58.67  &     3.57  &  0.151  &  0.898  &  1000 \\ 
1  &  1.0  &  1536(4)  &  1728(3)  &    480.7  &    3.416  &    77.45  &    62.59 &   45.62  &     4.93  &  0.195  &  0.962  &  1000 \\ 
3  &  1.0  &  1536(4)  &  1536(3)  &    421.2  &    1.709  &    50.42  &    42.70 &   26.80  &     10.2  &  0.295  &  0.997  &  1600 \\ 
10  &  1.0  &  1024(4)  &  1536(2)  &    313.1  &    1.166  &    31.38  &    27.77 &   14.62  &     21.9  &  0.458  &  1.002  &  4000 \\ 
30  &  0.5  &   768(3)  &  1536(2)  &    135.1  &    1.025  &    15.52  &    14.10 &   6.502  &     41.1  &  0.731  &  1.002  &  4500 \\ 

\end{tabular}
\caption{Numerical details and key responses for the group of cases with $Ra_S=10^{12}$.}
\label{tab:ras12}
\end{center}
\end{table}

\bibliographystyle{jfm}

\end{document}